\documentclass[aps,pra,twocolumn,superscriptaddress,lengthcheck,showpacs,amsmath,amssymb,floatfix,longbibliography]{revtex4-1}
\usepackage{graphicx}  
\usepackage{microtype}
\usepackage{float}
\usepackage{blindtext}
\usepackage{bbold}
\usepackage{multirow}
\usepackage{ORI_Group_style}

\hypersetup{colorlinks,linkcolor=blue,citecolor=blue,urlcolor=blue}

\begin{document}
\renewcommand\thesubsection{\arabic{subsection}}

\title{Quantum Motional State Tomography with Non-Quadratic Potentials and Neural Networks}

\author{Talitha Weiss}
\affiliation{Institute for Quantum Optics and Quantum Information of the Austrian Academy of Sciences, A-6020 Innsbruck, Austria}
\affiliation{Institute for Theoretical Physics, University of Innsbruck, A-6020 Innsbruck, Austria}

\author{Oriol Romero-Isart}
\affiliation{Institute for Quantum Optics and Quantum Information of the Austrian Academy of Sciences, A-6020 Innsbruck, Austria}
\affiliation{Institute for Theoretical Physics, University of Innsbruck, A-6020 Innsbruck, Austria}


\begin{abstract}
We propose to use the complex quantum dynamics of a massive particle in a non-quadratic potential to reconstruct an initial unknown motional quantum state. We theoretically show that the reconstruction can be efficiently done by measuring the mean value and the variance of the position quantum operator at different instances of time in a quartic potential. We train a neural network to successfully solve this hard regression problem. We discuss the experimental feasibility of the method by analyzing the impact of decoherence and uncertainties in the potential.
\end{abstract}

\maketitle

\section{Introduction}

One of the most fascinating possibilities in quantum physics is to prepare the motional degrees of freedom of a massive particle in a quantum state. The non-classical features of such a state can be demonstrated by reconstructing its quantum density-matrix operator and showing that its associated Wigner function has negative values. Such an endeavor has been successfully achieved with ions, see Ref.~\cite{Leibfried_IonReview_2003} and references therein. Today, the field of quantum nano- and micromechanics aims to do the same with objects much more massive~\cite{Aspelmeyer_ReviewOM_2014}, for instance nano- and micro-particles, which contain billions of atoms~\cite{Romero_Isart_SuperposLiving_2010,Chang_levitatedNano_2010}. Such an exciting goal has many challenges, and a crucial one is the faithful reconstruction (also called quantum tomography) of the quantum motional state. 

Standard strategies to perform quantum motional state tomography~\cite{Vanner_OM_QST_2015} are to couple the motion of the particle to a few-level system~\cite{Wallentowitz_QST_ion_1995,Singh_WignerTomo_usingAtom_2010}, to transfer the mechanical state to a cavity electromagnetic mode whose state can be reconstructed with homodyne tomography~\cite{Parkins_QST_stateTransfer_1999}, to apply coherent displacements and phonon number measurements on the motional degree of freedom~\cite{Wallentwoitz_QSTunbalancedHomodyne_1996,Banaszek_QSTphotonCountin_1996,Poyatos_QST_ion_1996}, as done with ions see, \eg,~Ref.~\cite{Leibfried_expQSTion_1996}, or to measure the full position distribution function (\ie having access to all moments) at different instances of time in a harmonic potential \cite{Vanner_CoolingByMsmt_QST_2013,Romero-Isart_LevTheoryProtocols_2011}. In this article, we propose an alternative approach based on exploiting two distinctive features of levitated particles: (i) their low level of motional decoherence and (ii) the possibility to engineer the potential of the particle, in particular to let the particle coherently evolve in a non-quadratic potential. We show that by solely measuring the mean value and the variance of the position of the particle (\ie the first two moments only) as a function of time during the evolution in a non-quadratic potential, one can efficiently reconstruct the initial unknown quantum motional state (\eg~a given non-Gaussian state). Such reconstruction is a hard quantum regression problem that, as we show below, is ideally suited for neural networks.

This article is organized as follows: In Sec.~\ref{sec_model} we introduce the physical scenario and argue that the time evolution of the mean value and variance of the particle's position should allow us to reconstruct the initial state. In Sec.~\ref{sec_results} we present our results: First, in Sec.~\ref{subsec_protocol}, we explain the overall protocol and how we use and evaluate the neural network for quantum state tomography. Then, in Sec.~\ref{subsec_noDec}, we show quantum state tomography in the absence of decoherence. In Sec.~\ref{subsec_dec} we discuss the effects of decoherence on the achieved fidelity and in Sec.~\ref{subsec_realistic} we consider realistic scenarios where we take experimental limitations into account. In Sec. \ref{sec_summary} we summarize our results. Additional information can be found in the appendices: First, in Appendix \ref{sec_expansion}, we discuss an approximative analytical approach and how it becomes unfeasible in the regime relevant for quantum state tomography. Finally, in Appendix \ref{sec_NN}, we present technical details about the architecture and training of the neural networks.
	
\section{Model}\label{sec_model}

Let us consider the one-dimensional motion of a particle of mass $m$ in a quartic potential such that its coherent dynamics is described by the Hamiltonian
\begin{equation}\label{eq_Hamiltonian}
    \Hop = \frac{\Pop^2}{2m}+\lambda \Xop^4 = \hbar \omega_0\left( \frac{\pop^2}{4}+\frac{\xop^4}{\alpha^4}\right).
\end{equation}
Here, $\Xop=\xop x_0$ and $\Pop=\pop p_0$, with $\coms{\Xop}{\Pop}=\im \hbar$, are the position and momentum, and $\lambda$ the strength of the quartic potential. We have extracted units using $x_0=\hbar/(2 p_0) =[\hbar/(2m\omega_0)]^{1/2}$ and defined the inverse quarticity parameter $\alpha=[\hbar\omega_0/(\lambda x_0^4)]^{1/4}$, with the motivation that we will consider initial motional quantum states assumed to be prepared in a harmonic potential of frequency $\w_0$. Additionally, we consider a standard source of decoherence for levitated particles \cite{Romero_Isart_QuantumSup_Collapse_2011,Joos_book_ClassicalWorld_QuTheory_2003,Schlosshauer_book_QtoCl_2007}  that is of position localization type (\eg~due to recoil heating or a fluctuating white-noise force). In total, the evolution of the density-matrix operator of the motional state is described by the master equation

\begin{equation}\label{meq}
    \dot\rhoop = -\frac{i}{\hbar} \coms{\Hop}{\rhoop}- \Gamma  \coms{\xop}{\coms{\xop}{\rhoop}} = \hat{L} \rhoop.
\end{equation}
Here, $\Gamma$ is the decoherence rate. 

At $t=0$, the particle is assumed to be in an unknown motional quantum state $\hat \eta \equiv \hat \rho(0)$ that we aim to reconstruct. The system evolves according to \eqnref{meq} and the state at a later time $t \geq 0$ is given by $\rhoop(t)= \exp (\hat{L} t) \etaop$. The position $\Xop$ is sufficiently measured at different instances of time to retrieve the mean value and its variance, that is, to obtain the dimensionless trajectories $u_1(t) \equiv \avg{\xop(t)}- \avg{\xop(0)}= \tr \spares{ \xop \pares{\hat \rho(t)-\hat \eta}}$ and $u_2(t) \equiv \expect{\xop^2(t)}-\expect{\xop(t)}^2 = \tr \spares{\xop^2 \rhoop (t)} - \pares{\tr \spares{\xop \rhoop}}^2 $. Note that, as defined, $u_1(0)=0$, and hence one does not need to assume an absolute position measurement. In Appendix \ref{sec_input}, we show examples of the trajectories $u_1(t)$ and $u_2(t)$ numerically calculated by solving  \eqnref{meq} on a truncated Hilbert space using the python toolbox QuTiP~\cite{Johansson_qutip_2012,Johansson_qutip_2013}. The question addressed in this article is the following: Can the information provided by $u_1(t)$ and $u_2(t)$ be used to reconstruct the unknown quantum motional state $\hat \eta$?

The reconstruction of a quantum motional state $\hat \eta$ could be performed should one have full knowledge of the mean values $\avg{\xop}$ and $\avg{\pop}$ and all the moments defined by $G^{a,b} = \langle (\pop-\expect\pop)^a (\xop-\expect{\xop})^b\rangle_\text{Weyl}$, where $\avg{\cdot}_\text{Weyl}$ denotes the mean value of the Weyl-ordered product of operators calculated for the state $\hat \eta$, and $a,b$ are non-negative integers.  As we show in detail in Appendix \ref{sec_expansion} based on Refs.~\cite{Ballentine_MomentEquations_1998, Brizuela_PRD_2014,Brizuela_PRD_generalizedUncertainty_2014}, the trajectories $u_1(t)$ and $u_2(t)$ depend, for $t$ larger than a given critical time, on basically all the moments $G^{a,b}$ of the state at $t=0$. This is a manifestation of the non-linear quantum dynamics induced by the quartic potential and has two consequences. First, it shows that, indeed, the trajectories $u_1(t)$ and $u_2(t)$ should provide sufficient information to reconstruct $\hat \eta$. This is in contrast to a harmonic potential where $u_1(t)$ and $u_2(t)$ would depend at most on quadratic moments. Second, it shows that in a quartic potential, consequently, it is not possible to correctly approximate $u_1(t)$ and $u_2(t)$ as a function of a finite set of initial moments and, hence,  the regression problem of deriving $\hat \eta$ based on $u_1(t)$ and $u_2(t)$ is a hard problem that, to our knowledge, cannot be solved with analytical tools. Nevertheless, this problem is very well suited to a neural network trained by supervised learning. The neural network will not require us to input how exactly the initial moments affect the trajectories. Instead, the neural network will, based on the training examples, find by itself an internal representation of the underlying regression problem. We remark that such a setting, inferring the quantum state from the time evolution of observables, is very different from recent works using neural networks for quantum state tomography of systems of many qubits~\cite{Torlai_NN_QST_2018, Xu_NNforQST_2018,Xin_LocalMsmt_QST_NN_2018,Quek_adaptiveQST_NN_2018}, or for filtering experimental data before performing quantum state tomography~\cite{Palmieri_experimentalQSTenhanced_NN_2019}. 

\section{Results}\label{sec_results}

In this section we present our results on quantum state tomography based solely on trajectories in non-quadratic potentials. In Sec.~\ref{subsec_protocol} we explain the protocol and how we obtain our results using neural networks. In Sec.~\ref{subsec_noDec} we investigate the ideal decoherence-free scenario, while in \ref{subsec_dec} we include decoherence and give an estimate of the thereby introduced necessary conditions. Eventually, in Sec.~\ref{subsec_realistic}, we study realistic scenarios including experimental limitations.

\subsection{Protocol}\label{subsec_protocol}

Let us now introduce the overall procedure: We propose to train the neural network on simulated data and then use the trained network to deduce the initial quantum state from experimentally measured trajectories $u_1(t)$ and $u_2(t)$. Experimentally these trajectories could be obtained by repeatedly re-preparing a particle in the same initial state and then evolving it (in the absence of measurements) up to a time $t_1$ when the position is measured, for instance, via optical position detection \cite{Tebbenjohanns_optimalDetection_2019, Tebbenjohanns_sidebandAsym_2019}. Averaging over the many repetitions, this reveals the expectation value and variance of the position at this time $t_1$. Consecutively, this procedure is repeated evolving the particle (in the absence of a measurement) to a later time $t_1+\delta t$ in order to measure the next point of the trajectory. This is repeated for all points of the trajectory.
 	
We now explain how the neural network is trained and tested. Initial states $\etaop$ are randomly sampled from a Hilbert Schmidt ensemble of density matrices of dimensions $d\times d$. That is, we assume that $\etaop$ is prepared in a harmonic potential of frequency $\w_0$ with zero probability to contain more than $d-1$ excitations, an assumption motivated by experiments preparing non-Gaussian quantum states after the particle has been cooled near the ground state of a harmonic potential~\cite{Romero-Isart_LevTheoryProtocols_2011}. While such a subspace is considered for the initial state $\etaop$, note that during the evolution the state $\rhoop(t) = \exp(\hat L t) \etaop$ can populate a much larger space that is only limited by numerical restrictions in the integration of the master equation~\eqnref{meq}. With the input of the trajectories $u_1(t)$ and $u_2(t)$, the neural network reconstructs a density-matrix operator $\etaop_\text{est}$ of size $d\times d$ with infidelity
\begin{equation}
    1-F=1- \tr \spare{\sqrt{\sqrt{\etaop}\etaop_\text{est}\sqrt{\etaop}}}.
\end{equation}
We remark that, in practice, any prior knowledge about the initially prepared quantum state should be used to accordingly choose the subspace and sampling distribution of training states in order to optimize the performance. Here, we sample the trajectories with timesteps $\delta t=0.05/\omega_0$ and each data point is represented by a neuron in the input layer of the network. Throughout the article we used four hidden layers of $800$, $800$, $400$, and $200$ neurons and an output layer of $2d^2$ neurons, representing the real numbers defining $\etaop_\text{est}$. The output is interpreted as a complex matrix $M$ that, generally, does not strictly fulfill the conditions of a physical density-matrix operator (positive semi-definite with unit trace). Consequently, the reconstructed physical state is  obtained  via $\etaop_\text{est}={M^\dagger M}/\tr \spares{M^\dagger M}$~~\cite{Xu_NNforQST_2018}. The network is trained via supervised learning using the mean squared error as the loss function and a training set of $10000$ randomly drawn quantum states. All results shown in the figures are obtained from a validation set, \textit{i.e.}, another set of $10000$ random states that were not used during training. More details on the network architecture and training can be found in Appendix \ref{sec_NN}. 

\begin{figure} 
	\center
	\includegraphics[width=\columnwidth]{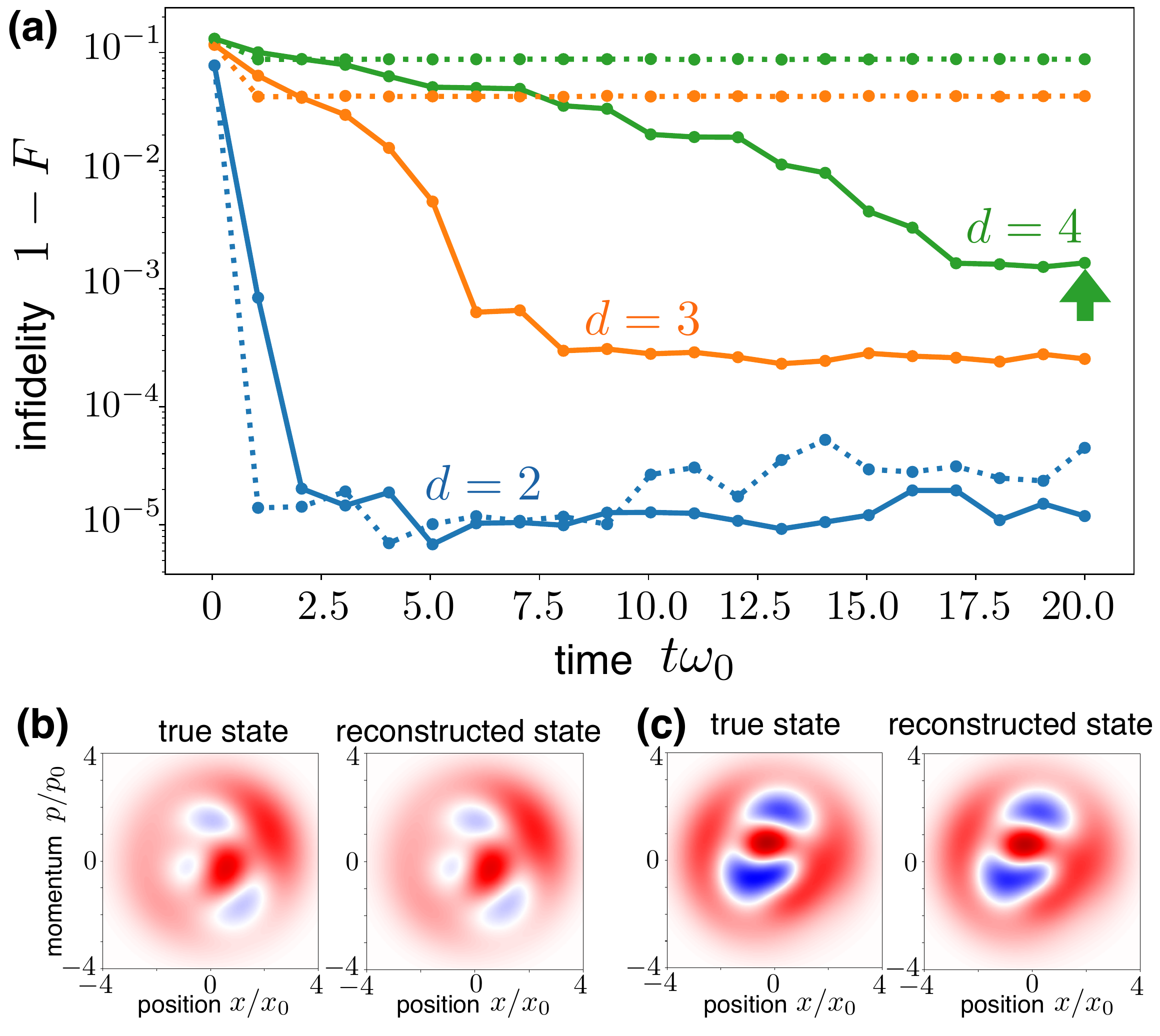}
\vspace{-0.2cm}
	\caption{\textit{Infidelity depending on trajectory length.} (a) Average infidelity $1-F$ of the quantum state predicted by the neural network as a function of the trajectory length that is used as an input. Each point presents a newly trained network using trajectories in a quartic (solid lines) or harmonic potential (dotted lines). The color denotes the dimension $d$ of the quantum state. (b) and (c) show the Wigner functions (red positive, blue negative values) of actual and reconstructed quantum state for two examples taken from the scenario indicated by the arrow in (a). The quantum state from the validation set where the neural network achieves the lowest (highest) infidelity $1-F$ is shown in (b), $1-F=4.8\times 10^{-5}$ ((c), $1-F=2.1\times 10^{-2}$). Parameters: $\alpha=5$, $\Gamma=0$. }
	\label{fig_noDec}
\end{figure}

\subsection{Decoherence-free scenario}\label{subsec_noDec}
Let us first show the results obtained in the absence of decoherence using \eqnref{meq} with $\Gamma=0$. In Fig.~\ref{fig_noDec}a, we show the average infidelity on the validation set reached by a neural network given input trajectories of a certain length (denoted by the time $t$) for $d=2,3$ and $4$, and with a quartic potential defined by the inverse quarticity $\alpha=5$. In every data point a new network was trained on the specific trajectory length (defining the input layer size of the network) and initial state dimension $d$ (defining the output layer size). In all cases, the infidelity decreases significantly with trajectory length and eventually saturates (all fluctuations around the saturation level are not of physical origin, as explained in Appendix \ref{sec_training}). The saturation occurs later for larger $d$, as an increasing number of $d^2-1$ independent moments need to be extracted from the trajectories in order to determine an arbitrary state of dimension $d$. The achieved infidelity also saturates on different levels depending on $d$, as we use the same number of training states ($10000$) despite the increasing size of the initial subspace. The achieved low infidelities demonstrate that the neural network can reconstruct the initial state from the trajectories with high accuracy, see Figs.~\ref{fig_noDec}b,c and caption for some examples. To show that the non-quadratic potential is indeed crucial, we also plot the performance of neural networks that were trained on trajectories in a harmonic potential (dotted lines in Fig.~\ref{fig_noDec}a) described by the Hamiltonian $\Hop = \Pop^2/(2m) + m \w_0^2 \Xop^2/2$. As expected, the non-quadratic potential outperforms the quadratic one, with the exception of the $d=2$ case, where the trajectories in the quadratic potential contain information about the only three moments that are required to fully determine the state. 

\begin{figure} 
	\center
	\includegraphics[width=\columnwidth]{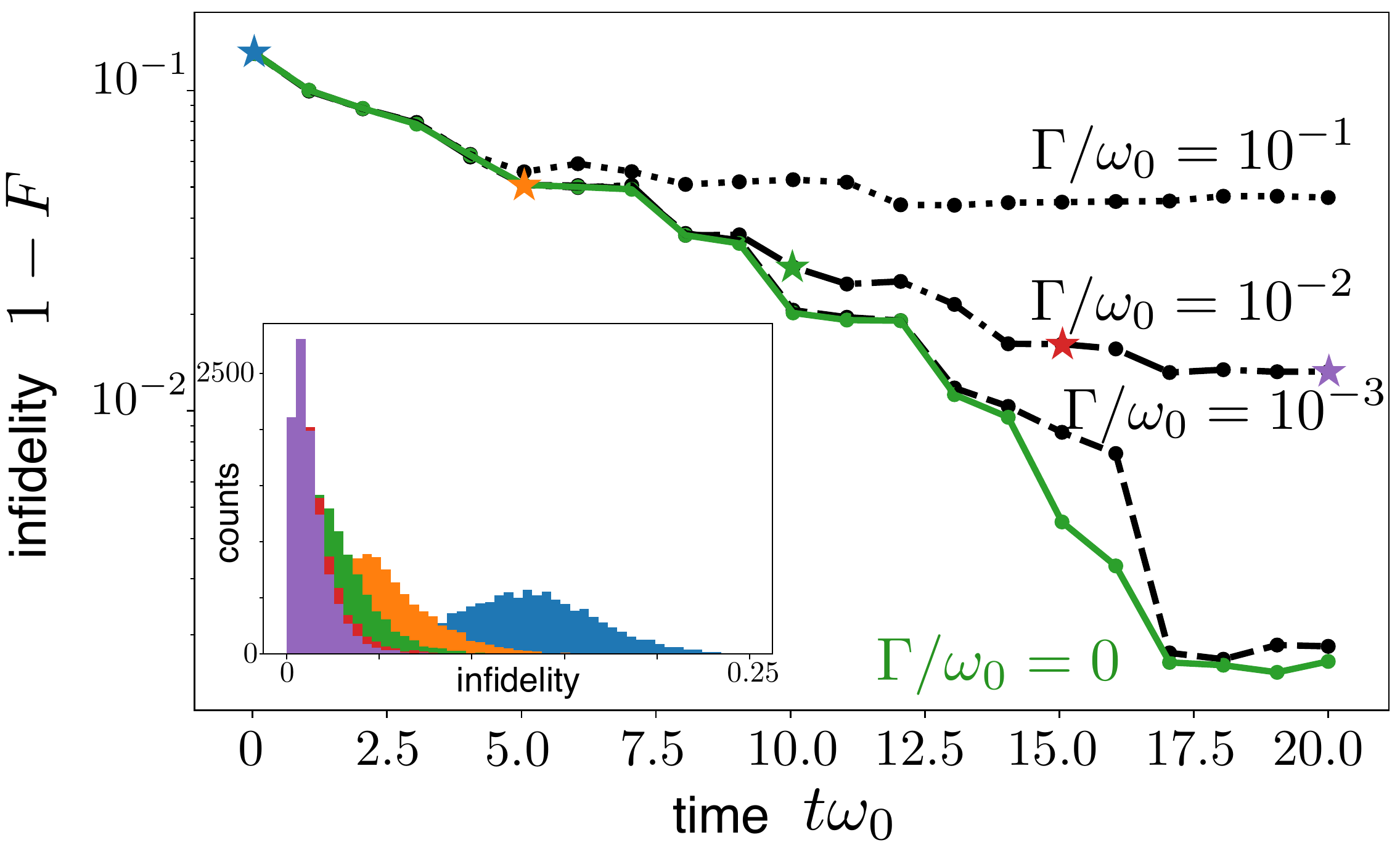}
\vspace{-0.2cm}
	\caption{\textit{Impact of decoherence on infidelity.} Infidelity $1-F$ as a function of time (as in Fig.~\ref{fig_noDec}a) for $d=4$ and $\alpha=5$ but including decoherence (black lines). For reference, the solid green line, same as in Fig.~\ref{fig_noDec}a, is the case without decoherence. The colored stars indicate the points for which the respective distribution of infidelities of all quantum states in the validation set is shown in the inset.  With longer trajectories the overall performance increases by improving both on the peak infidelity and the infidelity spread.  }
	\label{fig_wDec}
\end{figure}

\subsection{Limitations due to decoherence}\label{subsec_dec}
Let us now show the impact of decoherence, which will limit the length of the trajectories that can be used for quantum state reconstruction since, eventually, all information about the initial state is lost. In Fig.~\ref{fig_wDec}, we plot the achieved infidelities using \eqnref{meq} for different values of  $\Gamma$, an inverse quarticity $\alpha=5$,  the same set of quantum states as sampled for $d=4$ in Fig.~\ref{fig_noDec}a, and with neural networks trained and validated using the simulated trajectories in the presences of decoherence. The inset shows the distribution of infidelities reached in the validation set (see the caption for details). If the decoherence is sufficiently small (dashed line) the reached infidelity does not differ significantly from the performance achievable in absence of decoherence (green line, same as in  Fig.~\ref{fig_noDec}a), since the trajectories are only significantly altered by decoherence after  all information necessary to reconstruct the initial state has already been extracted. In contrast, at larger decoherence rates (dashed-dotted and dotted line), the trajectories are altered much earlier and both the average performance at intermediate times and the final performance become worse. The reason is that there is neither enough information contained in a trajectory up to the time where decoherence acts, nor sufficient time for the neural network to infer all the moments determining the initial state before decoherence erases the initial state dependence. 

The above discussion shows that in an experimental implementation of the proposed method the decoherence rate plays a limiting role. In the following we will show that eventually the ratio between decoherence and the strength of the non-quadratic potential is decisive. To this end, let us obtain a rough quantitative estimate of a necessary requirement. The initial state $\etaop$ is spatially confined in a length scale given by $ (\tr[\etaop \Xop^2])^{1/2} \sim x_0$ and has a kinetic energy of the order of $\hbar \omega_0$. During the evolution in the quartic potential, the initial state spreads as $ (\tr[\rho(t) \Xop^2])^{1/2} \sim x_0 \w_0 t$ and the effect of the quartic potential is relevant when the potential energy is comparable to the initial kinetic energy, namely for $t^\star$ such that $ \lambda (\tr[\rho(t^\star) \Xop^2])^{2} \sim \lambda (x_0 \w_0 t^\star)^4 \approx \hbar \omega_0$. Using the definition of the dimensionless inverse quarticity $\alpha=[4m^2\omega_0^3/(\hbar\lambda)]^{1/4}$, one obtains $t^\star \approx \alpha/\w_0$. Trajectories longer than $t^\star$ are required to be affected from the quartic potential, and the requirement that they are coherent demands that $1/\Gamma \gg t^\star$, or alternatively, $ \alpha  \ll \omega_0/\Gamma$. Thus, a stronger quartic potential (smaller inverse quarticity $\alpha$) helps to cope with decoherence. Such a rough estimate is a necessary but not a sufficient requirement, as can be seen in Fig.~\ref{fig_wDec}. Nevertheless, it provides a good  reference for experimental implementations. For instance, let us assume that the quartic potential is engineered using a potential of the form $V_G(X) = - m \w_0^2 \sigma^2 \exp [-X^2/\sigma^2]/2$ (\eg~generated by optical tweezers), which has been used to generate the quadratic potential, $V_G(X) \approx V_G(0) + m \w_0^2 X^2/2$. A (to leading order) pure quartic potential can then be obtained around $X=0$  by superimposing two such Gaussians $V(X) = V_G(X-\sigma/\sqrt{2})+  V_G(X+\sigma/\sqrt{2}) \approx 2 V_G(0) + \lambda X^4$, where $\lambda = m \w_0^2/[3 \sqrt{e} \sigma^2]$, and, hence, $\alpha \approx 1.8 \sqrt{\sigma/x_0}$. In this case, the necessary requirement reads $1.8 \sqrt{\sigma/x_0} \ll \omega_0/\Gamma$. For ions, this condition is not challenging  since $\w_0/\Gamma \sim 10^3$~\cite{Leibfried_IonReview_2003} and $x_0 \approx 10~\text{nm}$, and, hence, one requires potentials with $\sigma \ll 10^3~\mu\text{m}$. For optically levitated nanoparticles, where  $\w_0/\Gamma \sim 10^2$ \cite{Chang_levitatedNano_2010,Windey_Cooling_2019,Gonzalez-Ballestero_CoolingTheoery_2019}, and $x_0 \sim 10^{-12}~\text{m}$, the condition reads $\sigma \ll 10~\text{nm}$, which is not compatible with optical potentials where $\sigma$ is lower bounded by an optical wavelength. Therefore, levitated nanoparticles require either longer coherence times, achievable by evolution in the absence of recoil heating from laser light (quasi-electrostatic traps~\cite{Home_ionsAnharmonicPot_2011, Kuhlicke_NV_quadrupoleTrap_2014,Alda_NanoparticlePaulTrap_2016,Delord_SpinEchosLevi_2018,Bykov_NanoparticlesPaulTrap_2019}, magnetic traps~\cite{Romero-Isart_MagneticLevi_2012,Slezak_MagneticLevi_2018,Prat-Camps_InertialForceSensor_2017}, or in free fall \cite{Romero-Isart_LargeQuSup_2011,Hebestreit_FreeFallingNanoparticles_2018} where the quartic potential is only applied after the state has sufficiently broadened) or the use of electromagnetic forces near surfaces \cite{Diehl_NanoparticleCloseToMembrane_2018,Magrini_NearFieldNanoparticle_2018,Pino_Skatepark_2018} such that $\sigma$ can be potentially smaller than an optical wavelength. Instead of aiming for stronger non-quadratic potentials or longer decoherence times one could also speed up the broadening of the initially prepared state by introducing an inverted harmonic potential \cite{Pino_Skatepark_2018,ORI_CoherentInflation_2017} at the center of the quartic trap, that is, using a double-well potential.

\begin{figure}
	\center
	\includegraphics[width=\columnwidth]{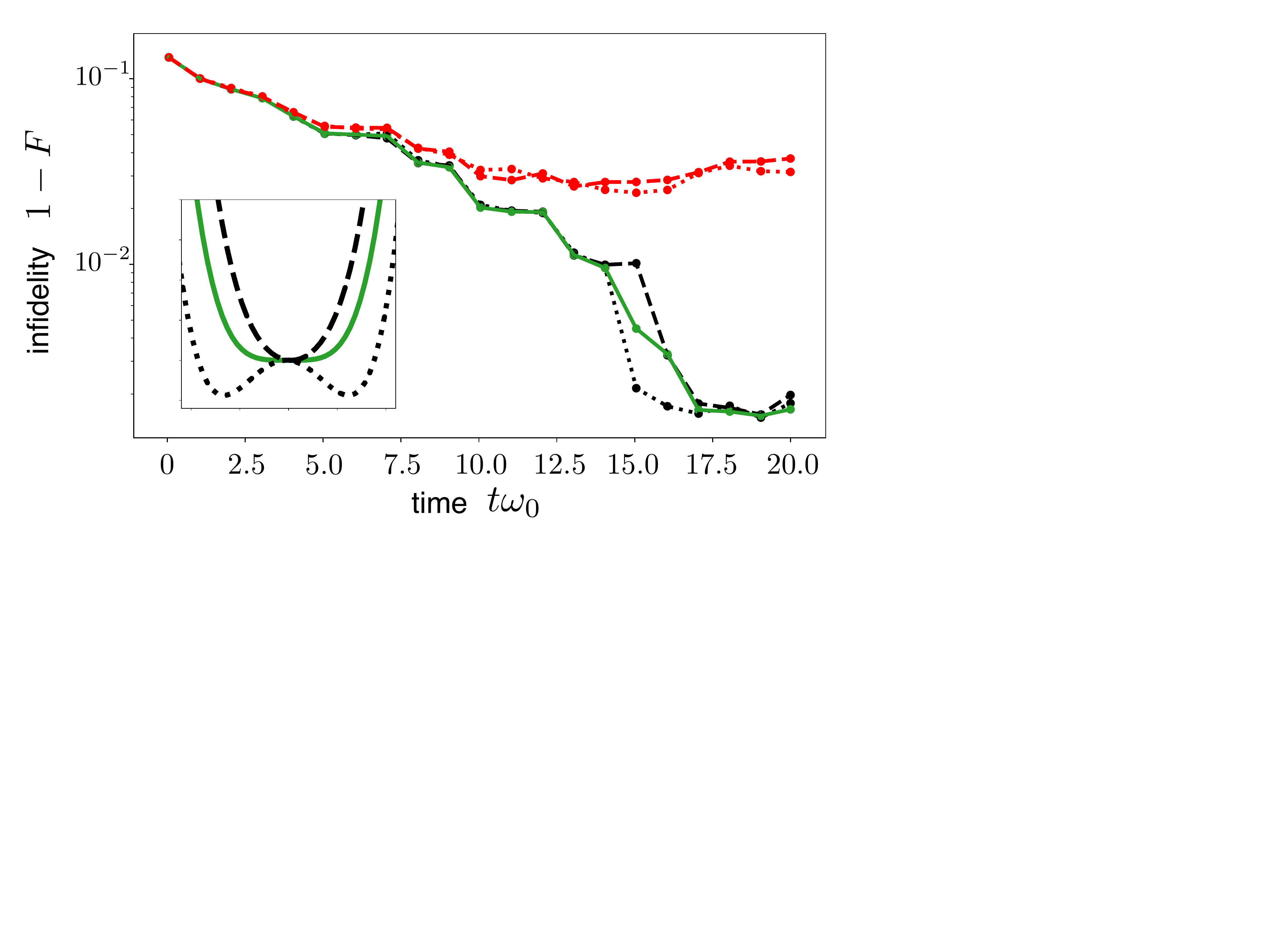}
\vspace{-0.2cm}
	\caption{\textit{Imperfect quartic potentials.} Infidelity $1-F$ as a function of time (as in Fig.~\ref{fig_noDec}a) for $\Gamma=0$, $d=4$, and $\alpha=5$. The solid green line shows the results in a perfect quartic potential (same as in Fig.~\ref{fig_noDec} (a)), while the dashed and dotted lines show the performance for the perturbed potentials, $\epsilon/x_0 =-0.1$ and $\epsilon/x_0=0.1$ respectively, and with $\alpha=5=\sqrt{\sigma/x_0}(6\sqrt{e})^{1/4}$. The black lines show the infidelity reached by neural networks both trained and validated on the perturbed potential.
	The red lines show the infidelity reached by neural networks trained on the purely quartic potential but applied to trajectories from the perturbed potential. The inset illustrate the respective potential shapes.}
	\label{fig_wDisp}
\end{figure}

\subsection{Realistic scenarios}\label{subsec_realistic}
Regarding the experimental implementation of the method, it is also clear that a perfect quartic potential cannot be engineered. Related to the discussion above, let us assume that the two Gaussian potentials are not perfectly symmetric aligned, namely one has $V_\epsilon (X) = V_G(X-\sigma/\sqrt{2}) + V_G(X+\epsilon+\sigma/\sqrt{2})$, where $\epsilon$ parametrizes the imperfection of the quartic potential. The form of such imperfect potentials is illustrated in the inset of Fig.~\ref{fig_wDisp} (dotted line for $\epsilon>0$, dashed line for $\epsilon<0$). In Fig.~\ref{fig_wDisp} we show the quantum state tomography performance of neural networks trained and tested on trajectories from the perturbed potential (black dotted line: $\epsilon/x_0=0.1$, black dashed line: $\epsilon/x_0=-0.1$).  A similar overall performance can be achieved compared to the purely quartic potential $\epsilon=0$ (solid green line). Thus, the neural network finds an appropriate model to each scenario and the quantum state tomography does not crucially depend on the details of the non-quadratic potential, even in the presence of small linear and quadratic contributions. 

So far, the training and testing scenario of the neural network were always the same. However, the experimental situation might not perfectly match the scenario used for training. For example, one could be ignorant of the exact form of the potential and hence use a neural network trained in slightly different potentials. The red lines in Fig.~\ref{fig_wDisp} show the reached average infidelity of neural networks that were trained on trajectories from the purely quartic potential ($\epsilon=0$) but that are then used to estimate the quantum state given trajectories from the perturbed potential (dotted and dashed again refer to $\epsilon/x_0=0.1$ and $\epsilon/x_0=-0.1$, respectively). At very short trajectory lengths the internal model of the trained neural network allows to reconstruct the quantum state with similar infidelity to the scenarios where training and validation situation had the same physical origin. If longer trajectories are used, the performance still improves, although the network is not able to retrieve as much information as in the ideal scenario. Given any specific accuracy goal, a numerical study would easily allow to estimate beforehand what size of experimental uncertainties a neural network trained with ignorance could bear.

\section{Summary}\label{sec_summary}
In summary, we have proposed a method to perform quantum motional state tomography for levitated particles (\eg~ions, nanoparticles), based on inferring the initial state from the time evolution of a few moments in a non-quadratic potential. The reconstruction is efficiently done with a neural network. We have analyzed the impact of decoherence and potential imperfections. As a proof-of-principle, we have shown results for a quartic potential in which the mean value and variance of the position is measured. We emphasize, however, that the method is very general since a neural network allows to optimally adapt the quantum state tomography to any given physical scenario in an experiment by using training examples from the particular situation. For the case of levitated nanoparticles, ground-state cooling is closely approached~\cite{Tebbenjohanns_optimalDetection_2019,Windey_Cooling_2019,Delic_Cooling_2019,Gonzalez-Ballestero_CoolingTheoery_2019, Tebbenjohanns_sidebandAsym_2019, Jain_LeviRecoilHeating_2016,Fonesca_NonlinearDynNanopart_2016,Millen_CoolingChargedNanosphere_2015,Asenbaum_CoolingSiliconNanopart_2013,Kiesel_CoolingSubmicronPart_2013,Gieseler_FeedbackCoolingNanopart_2012,Li_CoolingMicrosphere_2011}, hence the development of quantum tomography schemes is not only important but timely. At the same time, implementing non-quadratic potentials is also a fantastic tool to prepare non-Gaussian states. We therefore hope that this work will further motivate experimentalists in the field of levitated nanoparticles to engineer non-quadratic potentials to bring and probe nanoparticles in the quantum regime.

\begin{acknowledgements}
We acknowledge discussions with L. Novotny, F. Tebbenjohanns, and A. E. Rubio L\'opez. We thank the European Union Horizon 2020 Project FET-OPEN MaQSens (Grant No. 736943) for financial support. T.W.~also acknowledges financial support from the Alexander von Humboldt foundation.
\end{acknowledgements}

\appendix
\setcounter{equation}{0}
\renewcommand{\theequation}{A\arabic{equation}}
\section{\label{sec_expansion}Expansion in moments}

In this appendix, we discuss an approximative approach to analytically describe the time evolution of the initial quantum state and demonstrate how it becomes unfeasible in the regime important for quantum state tomography. This approach is based on truncating the infinite system of coupled equations of motion of all moments.

To this end, we consider the motion of a particle in a quartic trap as described by the Hamiltonian (\ref{eq_Hamiltonian}). Using the results of Refs.~\cite{Ballentine_MomentEquations_1998, Brizuela_PRD_2014,Brizuela_PRD_generalizedUncertainty_2014}, one can write the equations of motions of all moments:

\begin{align*}
	\langle\dot{\hat x}\rangle &= \expect{\pop}\\
	\langle\dot{\hat p}\rangle &= -\frac{8}{\alpha^4} \left (\expect\xop^3+3 \expect\xop G^{0,2}+G^{0,3}\right)\\
	\dot G^{a,b} =& b G^{a+1,b-1}+8 \frac{a}{\alpha^4} \left[ 3 \expect\xop G^{0,2}+G^{0,3}\right]G^{a-1,b}\\
	&-8 \frac{a}{\alpha^4}\left[3 \expect\xop^2G^{a-1,b+1}+3\expect\xop G^{a-1,b+2}+G^{a-1,b+3}\right]\\
	&+ 8 \frac{a}{\alpha^4}(a-1)(a-2)\left[\expect\xop G^{a-3,b}+G^{a-3,b+1} \right]
\end{align*}
Recall that $G^{a,b} = \langle (\pop-\expect\pop)^a (\xop-\expect{\xop})^b\rangle_\text{Weyl}$ denotes moments of the order $a+b$, with $\avg{\cdot}_\text{Weyl}$ the expectation value of the Weyl-ordered operators. This set of differential equations is exact but infinite. However, an approximation can be applied by only keeping all moments with combinations of $a$ and $b$ such that $a+b\leq N_\text{t}$, where $N_\text{t}$ is the truncation order. The resulting system of coupled, non-linear differential equations can then be solved numerically. 

\begin{figure*} 
	\center
	\includegraphics[width=\linewidth]{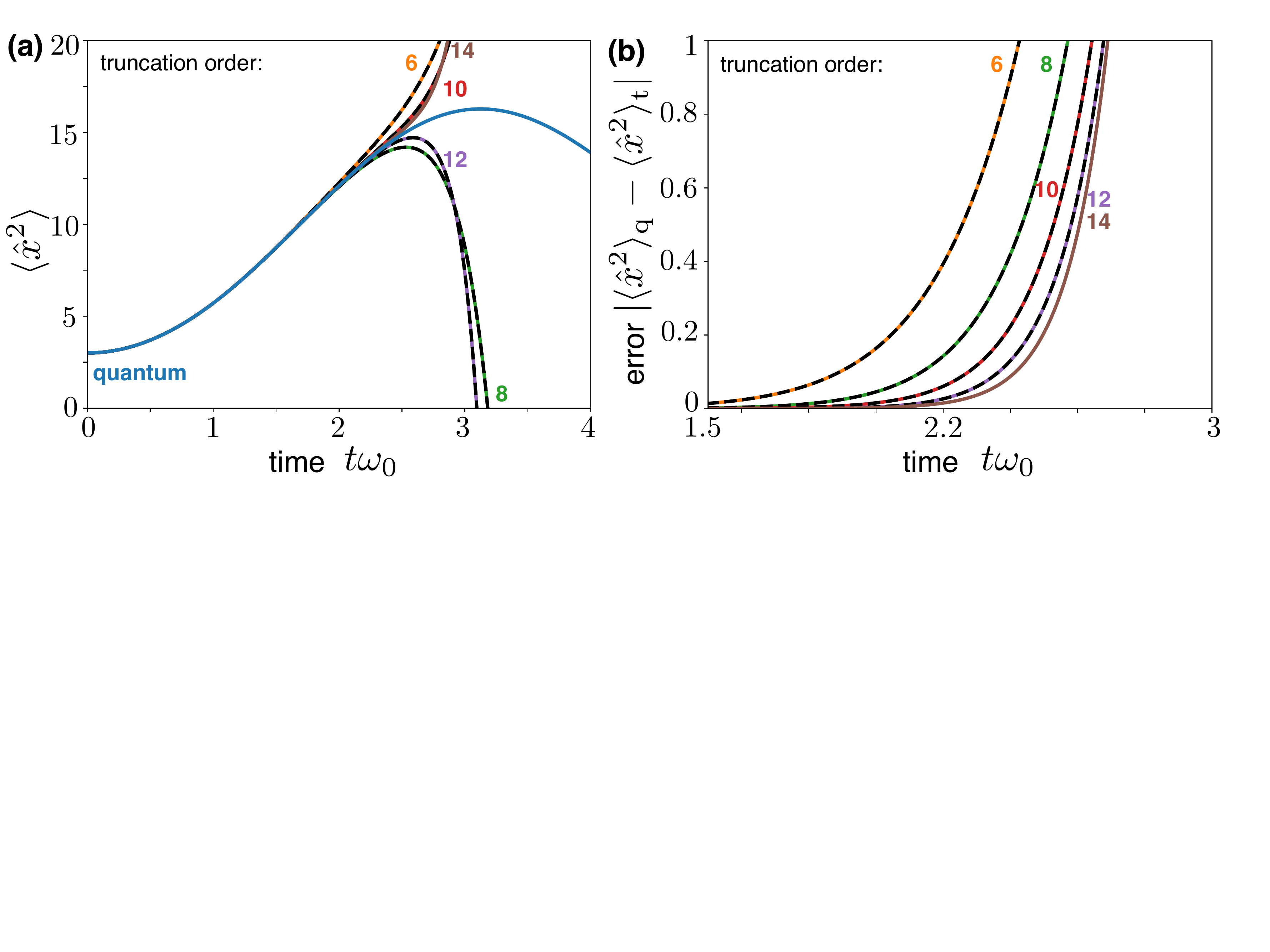}
	\vspace{-0.2cm}
	\caption{(a) Trajectory $\expect{\xop^2}$ of the initial Fock state $\ket{1}$ as obtained from a full quantum simulation (blue line) on a sufficiently larger Hilbert space, or from an expansion in moments truncated to the order indicated by the respectively colored number. The dashed black lines show the expansion in moments up to the next higher odd order, e.g.~up to order $7$ for the black dashed line on top of the yellow line which corresponds to truncation order $6$. For this particular initial state the odd order moments have zero or vanishing contribution to the overall motion. (b) shows the with time increasing absolute error between $\expect{\xop^2}_\text{t}$ from a truncated moment approach (truncation order again indicated by colored numbers) and the full quantum solution $\expect{\xop^2}_\text{q}$. Notably, the gain of accuracy by increasing the truncation order (by two) becomes smaller and smaller.}
	\label{fig_expansion}
\end{figure*}

In Fig.~\ref{fig_expansion} we illustrate the performance of this approximation using the Fock state $\ket{1}$ as the initial state, $\alpha=5$, and no decoherence $\Gamma=0$. The symmetry of a Fock state leads to vanishing first order moments, \ie, $\expect\xop=\expect\pop=0$ for all times, and odd order moments do not significantly contribute to the motion (the results of truncating to an odd order are shown as black dashed lines and coincide with the respective solution of the next lower even truncation order). In Fig.~\ref{fig_expansion}a we also display the exact solution (blue line) which we obtain by numerically integrating the Schr\"odinger equation using QuTiP.  We show approximations up to $N_\text{t}=14$. In Fig.~\ref{fig_expansion}b, we show the relative error between the full quantum solution $\expect{\xop^2}_\text{q}$ and the truncated  solution $\expect{\xop^2}_\text{t}$. Increasing the truncation order increases the time where the error remains below a certain threshold only slightly. Even more importantly, the gain of accuracy by increasing the truncation order (by two) decreases. Therefore, truncating the number of moments used to simulate the non-linear dynamics is not a good approximation. This means that the trajectories depend, in general, on an unbounded number of initial moments, and hence provide sufficient information to reconstruct the initial quantum state. Such task for arbitrary quantum states cannot, however, be done analytically but with a neural network, as we show in this work.

\setcounter{equation}{0}
\renewcommand{\theequation}{B\arabic{equation}}

\section{\label{sec_NN}Neural network}
In this appendix, we provide additional information on the neural networks that we used throughout this work. In particular, in Appendix \ref{sec_hyper}, we summarize the network architecture and relevant hyperparameters. In Appendix \ref{sec_input} we discuss the input to the neural network and show example trajectories for the various scenarios discussed in the main text. Finally, in Appendix \ref{sec_training}, we describe the technical aspects of training the neural network in detail and present a typical learning curve.

\subsection{\label{sec_hyper}Details on the neural network}
The neural networks used throughout this work were the same for all tasks, with only the input and output layer depending on the trajectory length and dimensionality of the specific problem respectively. In Table \ref{tbl_NN} we describe the network architecture and specify the relevant hyperparameters.

\begin{table*}
	\begin{center}
		\begin{tabular}{|| p{5cm} | p{10cm}||} 
			\hline
			\textbf{network type} & feed-forward, densely connected \\ 
			\hline
			\textbf{neurons} & \multirow{3}{10cm}{\textit{input layer:} $2N$ ($N$ the number of points in each trajectory)\\ \textit{hidden layers:} $800$, $800$, $400$, $200$\\ 
			\textit{output layer:} $2d^2$ ($d$ the dimension of the initial quantum state)} \\
			&\\
			&\\
			
			\hline
			\textbf{activation functions} & ‘sigmoid', for the final layer ‘tanh' \\
			\hline
			\textbf{optimizer} & adam  (with keras default values) \\ 
			\hline
			\textbf{loss function} & custom mean-squared error \\ 
			\hline
			\textbf{end of training} & early stopping, with a patience of $500$ epochs (maximal $10000$ epochs)\\ 
			\hline
			\textbf{batch size} & 512  \\
			\hline
			\textbf{number of training states} & 10000  \\
			\hline
			\textbf{number of validation states} & 10000  \\
			\hline
		\end{tabular}
	\caption{Summary of the neural network architecture and hyperparameters.}\label{tbl_NN}
	\end{center}
\end{table*}

\subsection{\label{sec_input}Input to the neural network}

\begin{figure*} 
	\center
	\includegraphics[width=\linewidth]{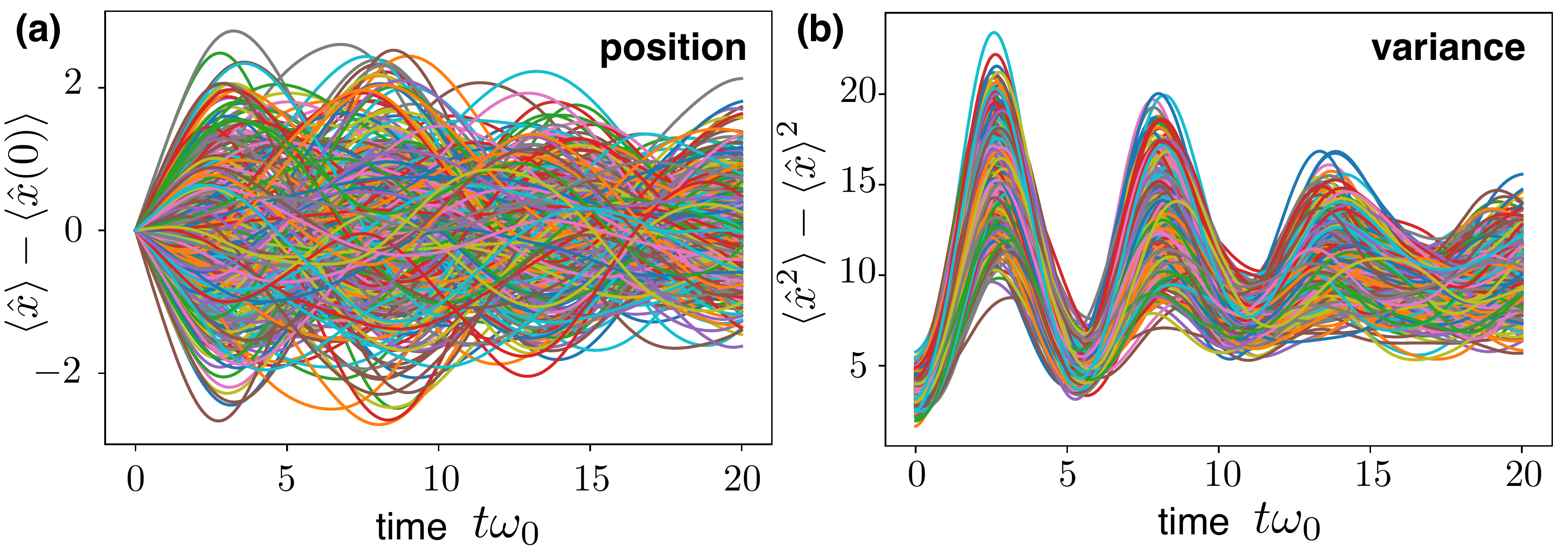}
	\vspace{-0.2cm}
	\caption{Four hundred example trajectories of shifted position, (a), and variance, (b), taken from the training set of four dimensional states, $d=4$, in a quartic potential. The neural network receives a single position trajectory and the corresponding single variance trajectory as an input. Parameters: $\alpha=5$, $\Gamma=0$.}
	\label{fig_trajs_noDec}
\end{figure*}

\begin{figure*} 
	\center
	\includegraphics[width=\linewidth]{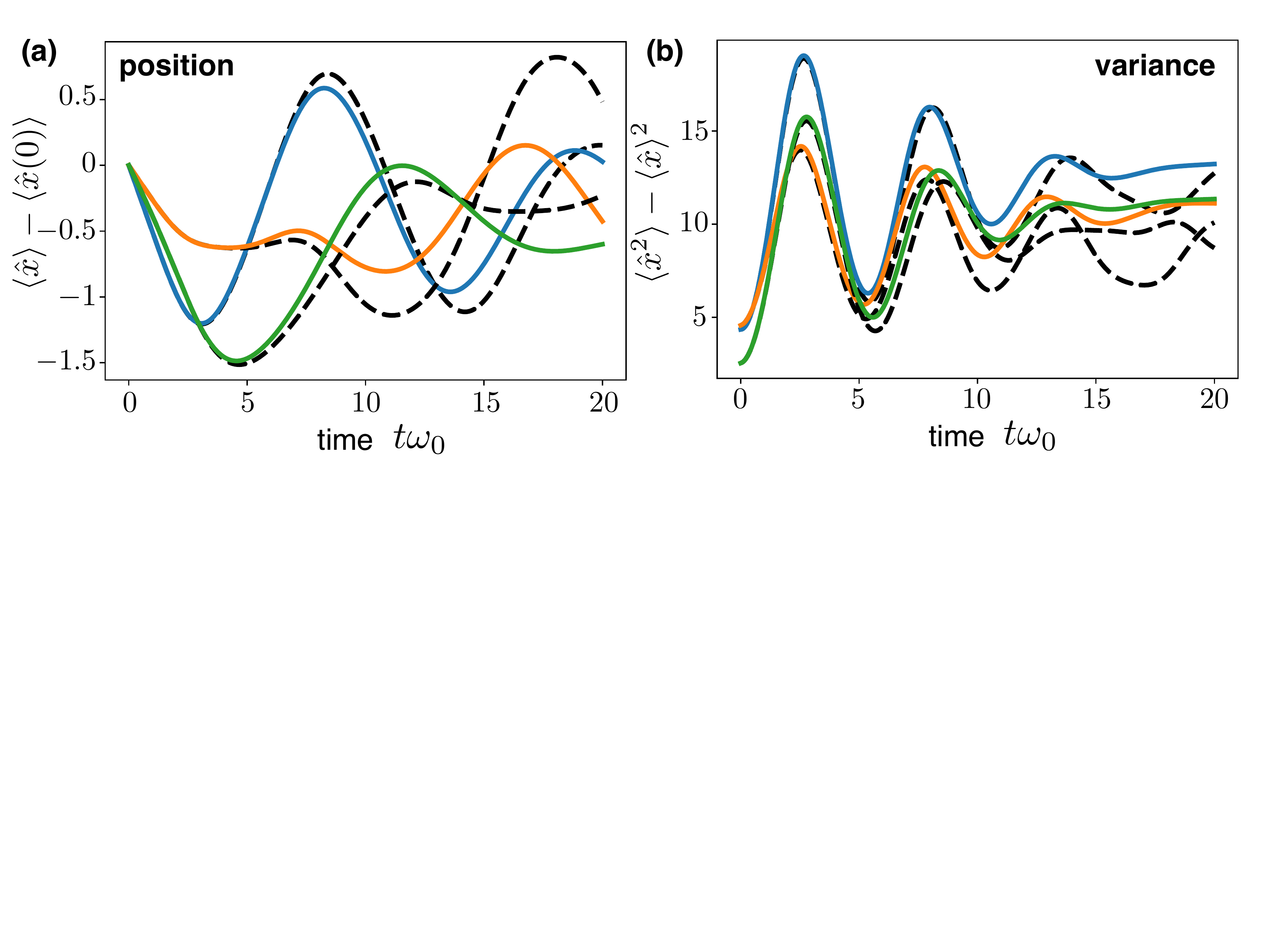}
	\vspace{-0.2cm}
	\caption{Three example trajectories of shifted position, (a), and variance, (b), taken from the training set of four dimensional states, $d=4$, in a quartic potential. The dashed black lines show the `original' decoherence-free trajectories. The colored lines show the respective trajectory of the same initial state in the presence of recoil heating $\Gamma/\omega_0=10^{-2}$. Parameters: $\alpha=5$.}
	\label{fig_trajs_wDec}
\end{figure*}

\begin{figure*} 
	\center
	\includegraphics[width=\linewidth]{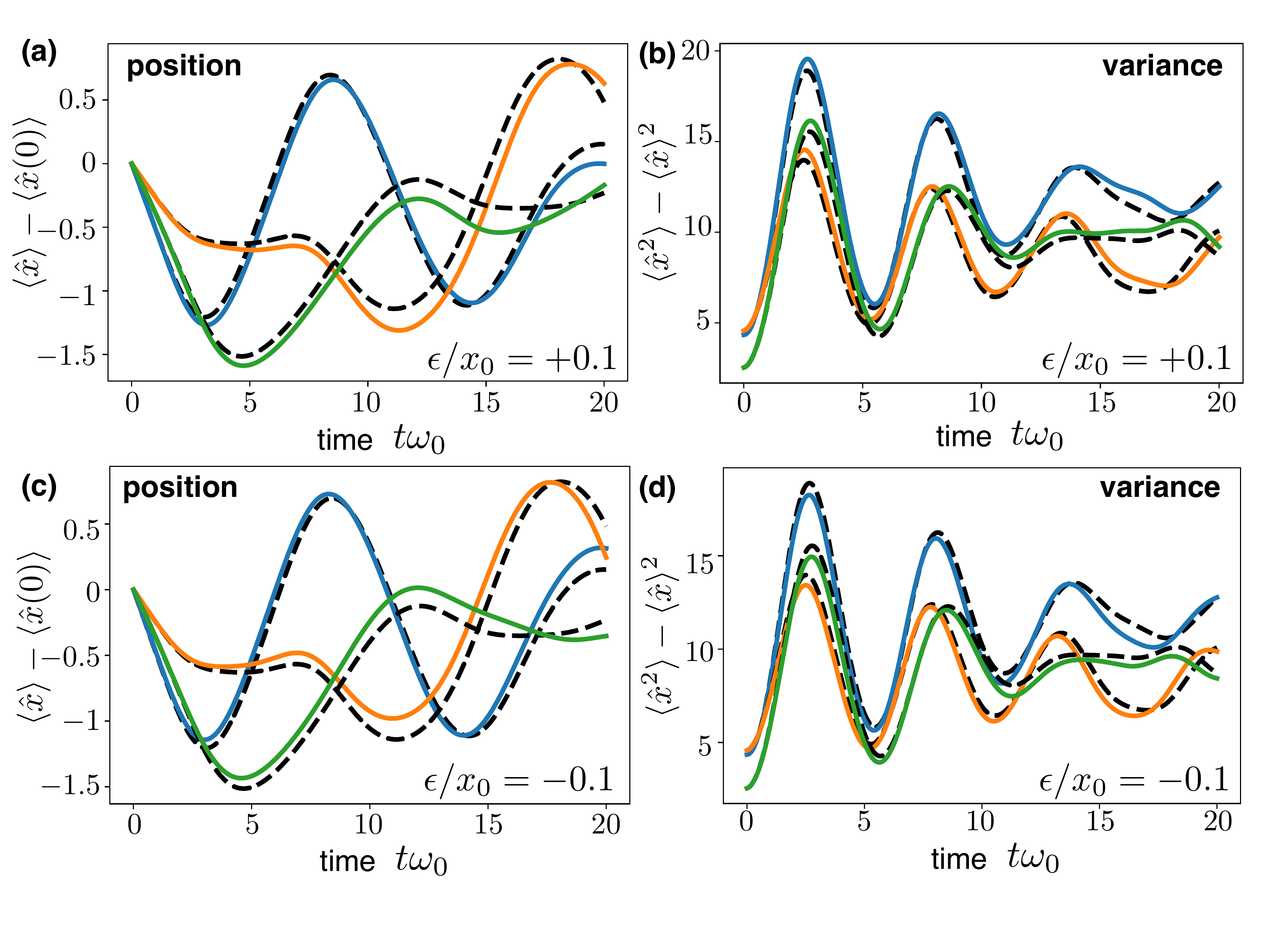}
	\vspace{-0.2cm}
	\caption{Three example trajectories of shifted position, (a) and (c), and variance, (b) and (d) taken from the training set of four dimensional states, $d=4$. The dashed black lines show the `original' trajectories in a purely quartic potential. The colored lines show the respective trajectory of the same initial state in the perturbed non-quadratic potential, i.e., $\epsilon/x_0=0.1$ for (a) and (b) and $\epsilon/x_0=-0.1$ for (c) and (d). Here, $\Gamma/\omega_0=0$ and $\alpha=5$ for all trajectories.}
	\label{fig_trajs_wDisp}
\end{figure*}

Here, we show examples of trajectories that were used for training the neural network to give an impression about their diversity, Fig.~\ref{fig_trajs_noDec}, the impact of decoherence, Fig.~\ref{fig_trajs_wDec}, and the motion in a perturbed non-quadratic potential, Fig.~\ref{fig_trajs_wDisp}. 

The effect of position localization type of decoherence, Fig.~\ref{fig_trajs_wDec}, reveals a damping of the position and eventually an increase in variance (after the oscillations disappeared). As expected, all trajectories approach each other at long times (even more significantly for times later than the ones shown and used throughout this work). This is a manifestation of the loss of information about the initial state.

In contrast, the deviation from a purely quartic potential does alter the trajectories in a systematic manner. The position and variance show no sign of damping or heating (in those simulations no decoherence was included, only a changed potential) and the trajectories remain distinguishable. Thus, the initial state dependence remains although the mapping from the trajectory to the state is changed. A neural network trained on either scenario will therefore end up with a different internal representation to reconstruct the quantum state in an optimal way.

For the input of the neural network, we sampled each trajectory of maximal length $t\omega_0=20$ with $400$ data points, resulting in a spacing of $\delta t\omega_0=0.05$. However, this is not crucial for the training of the network as long as the number of sampled data points is sufficiently high to accurately represent the trajectories. The data points from the (shifted) position and variance trajectory are concatenated to one large input vector that is then fed to the input layer of the neural network. For the figures in the main text that show the performance of the quantum state tomography with shorter trajectories we use the \textit{same} trajectories, but instead of using all $400$ data points as an input we shorten it appropriately. Thereby, we only had to simulate the trajectories once with full length instead of creating many different data sets for various lengths. 

\subsection{Training}\label{sec_training}

\textit{Loss function:}
The neural networks are trained via standard supervised training using keras. However, as discussed in the main text, the output of the neural network cannot directly be interpreted as the desired density matrix. Instead, we interpret the output vector of size $2d^2$ as the real and imaginary parts of the $d^2$ entries of a complex matrix $M$ from which the density matrix is obtained via $\etaop_\text{est}={M^\dagger M}/\tr [M^\dagger M]$. Therefore, we define a custom loss function where we first calculate the real and imaginary parts of all entries of the estimated density matrix $\etaop_\text{est}$. These values form  a vector $\vec{\eta}_\text{est}$ with $2d^2$ entries (with some values equal to zero by definition since the density matrix is Hermitian). Similarly, we write the true density matrix $\etaop$ as a vector $\vec{\eta}$ of length $2d^2$. Finally we calculate the element-wise mean-squared error between $\vec{\eta}$ and $\vec{\eta}_\text{est}$ which we use as the loss to train the neural network. 

\textit{Early stopping:}
Instead of training all neural networks for a fixed amount of epochs, we made use of an early stopping routine. Thereby training is stopped when no significant improvement is achieved anymore (or at a chosen maximum of $10000$ epochs). In particular, we have used a ‘patience' value of $500$, \ie if the so far lowest validation loss is not lowered even further (by the tiniest amount) within the following $500$ epochs then training is stopped. The neural network state achieving the lowest validation loss reached so far is stored and eventually used for quantum state reconstruction. The usage of early stopping prevents us from spending too much computational time if significant progress is no longer made (with the risk of missing very slow improvements) and is a common tool to avoid ending up in overfitting. Note that we attribute the fluctuations around (and small increase above) the saturation level shown in Fig.~1 of the main text (most prominently visible for the $d=2$ curves) to our particular choice of patience. Data points at larger $t\omega_0$ correspond to training a neural network with longer trajectories, \ie more input data are provided to a larger number of input neurons. However, all other parameters concerning the network architecture and training are kept fixed. It can be anticipated that this generically means that the overall learning rate is slowed down since a larger amount of data has to be processed with the same tools. This, in turn, makes it more likely that the next best value of validation loss is shifted out of the fixed patience interval. Randomly this might or might not be the case, leading to an earlier or later stop of the training and thereby to the fluctuations in performance (with a small trend of performing worse). This becomes particularly evident, if the trajectory is sufficiently long to gain all information (the regime where the performance saturates) and no huge further improvements can be expected anyways. Indeed, in this regime we notice that the stopping epochs of two consecutive data points (two networks that receive trajectories with just a small change in trajectory length) can differ significantly (\eg~one of the networks training almost twice as many epochs as the other). This effect can be avoided (or at least attenuated) by training all networks with the same, much larger, number of epochs while still storing the network with the overall best performance during training. However, this would significantly increase the overall computational time needed.

\begin{figure*}
	\center
	\includegraphics[width=\linewidth]{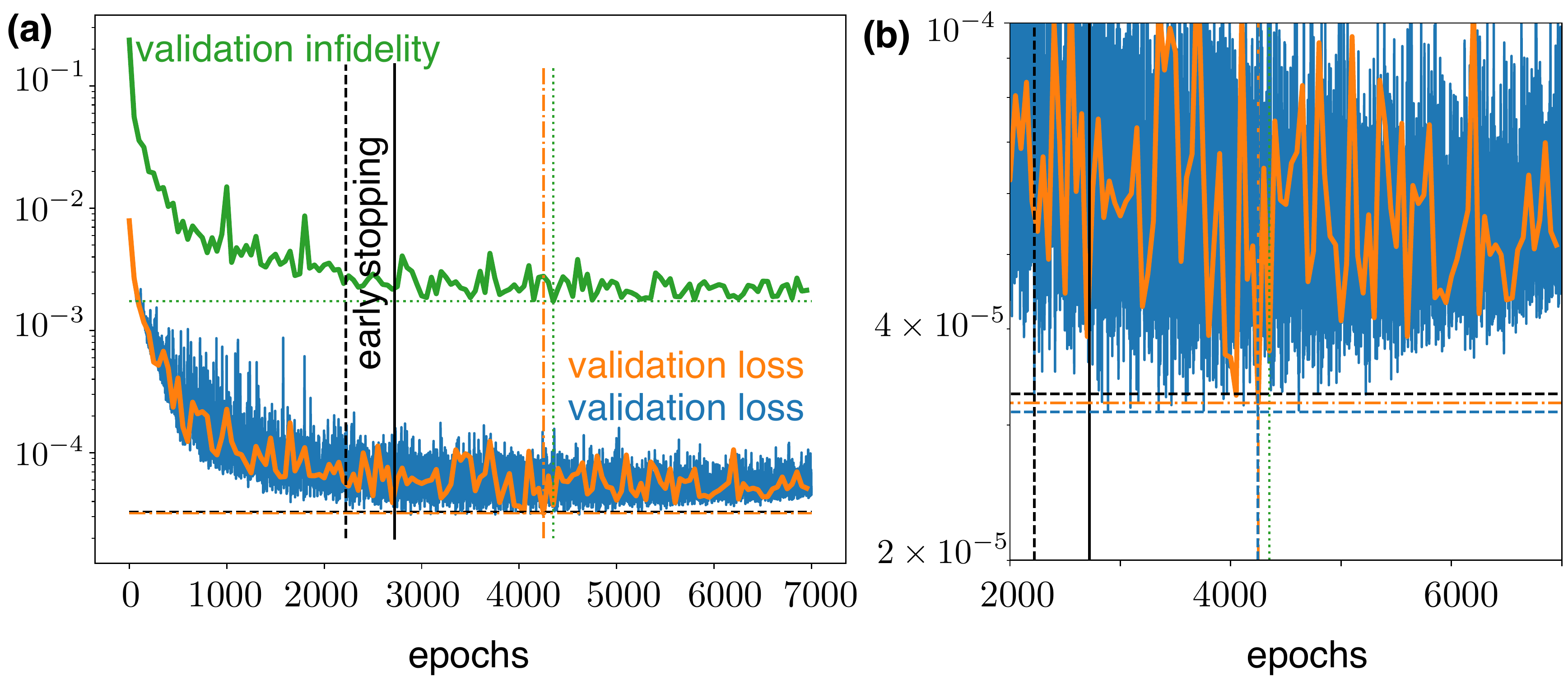}
	\vspace{-0.2cm}
	\caption{(a) Typical learning curve showing the performance of a neural network during training. The blue line shows the validation loss at every epoch. The orange line shows the validation loss only at every $50$th epoch where also the average infidelity of the validation set was computed (green line). The solid black line shows at which epoch the early stopping that we used for our main results would have stopped the presented training, with the black dashed line indicating the best performing network so far. The actual lowest value of validation loss can be seen in the zoom in (b) indicated by the blue dashed line and occurs at an even later epoch, close to where also the lowest value of infidelity is reached (green dotted lines). Physical scenario for training: perfect quartic potential, $d=4$, $\alpha=5$, $t\omega_0=20$, $\Gamma=0$.}
	\label{fig_learning_Curve}
\end{figure*}

\textit{Learning curve:}
In Fig.~\ref{fig_learning_Curve} we show a typical learning curve. In particular, we show the performance of a neural network that is trained to reconstruct four-dimensional quantum states from trajectories in the absence of decoherence and in a perfect quartic potential. The maximal trajectory length $t\omega_0=20$ was used. While the blue line shows the validation loss at every epoch, we only calculated the validation infidelity every $50$th epoch (green line) and the orange line  shows the validation loss at the same epochs. At very late epochs the validation loss rises again, indicating overfitting. For our main results, we used early stopping with a patience of $500$ epochs to avoid ending up in this regime. In Fig.~\ref{fig_learning_Curve}, the solid black line marks where the early stopping criterion would have canceled further training. The best validation loss value so far is indicated by the dashed black line which was not improved in the following $500$ epochs up to the solid black line. The actual lowest value of the validation loss in this figure occurred roughly $2000$ epochs later, as marked by the blue dashed line in the zoom in shown in Fig.~\ref{fig_learning_Curve}b. The minimum of the orange line, showing the validation loss on a rougher grid, occurs at a very similar epoch. Notably, the overall improvement of validation loss from the best value reached before the early stopping (black dashed line) and the actual best value (blue dashed line) is small. The improvement of the validation infidelity from the early stopping threshold to the actual lowest infidelity value indicated by the green dotted line is more significant though. The lowest validation loss (of the orange curve) and the lowest validation infidelity (green curve) do not coincide at the same epoch (although being relatively close). This shows that, although the mean squared error loss function works very well for training, it does not perfectly match the physical objective of small infidelity. Another interesting observation is that the infidelity does not immediately increase when the validation loss starts to rise again at very late training epochs shown in Fig.~\ref{fig_learning_Curve}.

Finally, we want to remark that throughout this work we used a total set of $20000$ randomly sampled quantum states, $10000$ each for training and validation. The states were sampled using QuTiP's {\tt rand\_dm\_ginibre()} function, where the default is to produce full rank matrices that correspond to sampling from a Hilbert Schmidt ensemble. For this fixed set of random states trajectories were simulated according to the discussed physical situations and with a maximal length of $t\omega_0=20$. Furthermore, the same fixed random seed was used for training all neural networks. Thereby, the comparison between our results, \eg~comparing the performance of neural networks on the same physical scenario but with different trajectory length, is simplified: The neural network trains on the \textit{same trajectories} in the \textit{same order}, just that those trajectories are a bit longer. The long trajectories contain the \textit{same information} as the short trajectories plus some additional data points that can either contribute additional information or not depending on the scenario. This is also a reason, why, \eg~in Fig.~1 of the main text, we can clearly state that the fluctuations around the saturation level (most prominently seen for both $d=2$ results) are not physical. Indeed, there cannot be less information in the longer trajectories than in the shorter ones and we have to attribute the observed deviations t o the neural network learning less efficiently.

\begin{thebibliography}{50}%
	\makeatletter
	\providecommand \@ifxundefined [1]{%
		\@ifx{#1\undefined}
	}%
	\providecommand \@ifnum [1]{%
		\ifnum #1\expandafter \@firstoftwo
		\else \expandafter \@secondoftwo
		\fi
	}%
	\providecommand \@ifx [1]{%
		\ifx #1\expandafter \@firstoftwo
		\else \expandafter \@secondoftwo
		\fi
	}%
	\providecommand \natexlab [1]{#1}%
	\providecommand \enquote  [1]{``#1''}%
	\providecommand \bibnamefont  [1]{#1}%
	\providecommand \bibfnamefont [1]{#1}%
	\providecommand \citenamefont [1]{#1}%
	\providecommand \href@noop [0]{\@secondoftwo}%
	\providecommand \href [0]{\begingroup \@sanitize@url \@href}%
	\providecommand \@href[1]{\@@startlink{#1}\@@href}%
	\providecommand \@@href[1]{\endgroup#1\@@endlink}%
	\providecommand \@sanitize@url [0]{\catcode `\\12\catcode `\$12\catcode
		`\&12\catcode `\#12\catcode `\^12\catcode `\_12\catcode `\%12\relax}%
	\providecommand \@@startlink[1]{}%
	\providecommand \@@endlink[0]{}%
	\providecommand \url  [0]{\begingroup\@sanitize@url \@url }%
	\providecommand \@url [1]{\endgroup\@href {#1}{\urlprefix }}%
	\providecommand \urlprefix  [0]{URL }%
	\providecommand \Eprint [0]{\href }%
	\providecommand \doibase [0]{http://dx.doi.org/}%
	\providecommand \selectlanguage [0]{\@gobble}%
	\providecommand \bibinfo  [0]{\@secondoftwo}%
	\providecommand \bibfield  [0]{\@secondoftwo}%
	\providecommand \translation [1]{[#1]}%
	\providecommand \BibitemOpen [0]{}%
	\providecommand \bibitemStop [0]{}%
	\providecommand \bibitemNoStop [0]{.\EOS\space}%
	\providecommand \EOS [0]{\spacefactor3000\relax}%
	\providecommand \BibitemShut  [1]{\csname bibitem#1\endcsname}%
	\let\auto@bib@innerbib\@empty
	\bibitem [{\citenamefont {Leibfried}\ \emph {et~al.}(2003)\citenamefont
		{Leibfried}, \citenamefont {Blatt}, \citenamefont {Monroe},\ and\
		\citenamefont {Wineland}}]{Leibfried_IonReview_2003}%
	\BibitemOpen
	\bibfield  {author} {\bibinfo {author} {\bibfnamefont {D.}~\bibnamefont
			{Leibfried}}, \bibinfo {author} {\bibfnamefont {R.}~\bibnamefont {Blatt}},
		\bibinfo {author} {\bibfnamefont {C.}~\bibnamefont {Monroe}}, \ and\ \bibinfo
		{author} {\bibfnamefont {D.}~\bibnamefont {Wineland}},\ }\bibfield  {title}
	{\enquote {\bibinfo {title} {Quantum dynamics of single trapped ions}},\
	}\href {\doibase 10.1103/RevModPhys.75.281} {\bibfield  {journal} {\bibinfo
			{journal} {Rev. Mod. Phys.}\ }\textbf {\bibinfo {volume} {75}},\ \bibinfo
		{pages} {281} (\bibinfo {year} {2003})}\BibitemShut {NoStop}%
	\bibitem [{\citenamefont {Aspelmeyer}\ \emph {et~al.}(2014)\citenamefont
		{Aspelmeyer}, \citenamefont {Kippenberg},\ and\ \citenamefont
		{Marquardt}}]{Aspelmeyer_ReviewOM_2014}%
	\BibitemOpen
	\bibfield  {author} {\bibinfo {author} {\bibfnamefont {M.}~\bibnamefont
			{Aspelmeyer}}, \bibinfo {author} {\bibfnamefont {T.~J.}\ \bibnamefont
			{Kippenberg}}, \ and\ \bibinfo {author} {\bibfnamefont {F.}~\bibnamefont
			{Marquardt}},\ }\bibfield  {title} {\enquote {\bibinfo {title} {Cavity
				optomechanics}},\ }\href {\doibase 10.1103/RevModPhys.86.1391} {\bibfield
		{journal} {\bibinfo  {journal} {Rev. Mod. Phys.}\ }\textbf {\bibinfo {volume}
			{86}},\ \bibinfo {pages} {1391} (\bibinfo {year} {2014})}\BibitemShut
	{NoStop}%
	\bibitem [{\citenamefont {Romero-Isart}\ \emph {et~al.}(2010)\citenamefont
		{Romero-Isart}, \citenamefont {Juan}, \citenamefont {Quidant},\ and\
		\citenamefont {Cirac}}]{Romero_Isart_SuperposLiving_2010}%
	\BibitemOpen
	\bibfield  {author} {\bibinfo {author} {\bibfnamefont {O.}~\bibnamefont
			{Romero-Isart}}, \bibinfo {author} {\bibfnamefont {M.~L.}\ \bibnamefont
			{Juan}}, \bibinfo {author} {\bibfnamefont {R.}~\bibnamefont {Quidant}}, \
		and\ \bibinfo {author} {\bibfnamefont {J.~C.}\ \bibnamefont {Cirac}},\
	}\bibfield  {title} {\enquote {\bibinfo {title} {Toward quantum superposition
				of living organisms}},\ }\href {\doibase 10.1088/1367-2630/12/3/033015}
	{\bibfield  {journal} {\bibinfo  {journal} {New J. Phys.}\ }\textbf {\bibinfo
			{volume} {12}},\ \bibinfo {pages} {033015} (\bibinfo {year}
		{2010})}\BibitemShut {NoStop}%
	\bibitem [{\citenamefont {Chang}\ \emph {et~al.}(2010)\citenamefont {Chang},
		\citenamefont {Regal}, \citenamefont {Papp}, \citenamefont {Wilson},
		\citenamefont {Ye}, \citenamefont {Painter}, \citenamefont {Kimble},\ and\
		\citenamefont {Zoller}}]{Chang_levitatedNano_2010}%
	\BibitemOpen
	\bibfield  {author} {\bibinfo {author} {\bibfnamefont {D.~E.}\ \bibnamefont
			{Chang}}, \bibinfo {author} {\bibfnamefont {C.~A.}\ \bibnamefont {Regal}},
		\bibinfo {author} {\bibfnamefont {S.~B.}\ \bibnamefont {Papp}}, \bibinfo
		{author} {\bibfnamefont {D.~J.}\ \bibnamefont {Wilson}}, \bibinfo {author}
		{\bibfnamefont {J.}~\bibnamefont {Ye}}, \bibinfo {author} {\bibfnamefont
			{O.}~\bibnamefont {Painter}}, \bibinfo {author} {\bibfnamefont {H.~J.}\
			\bibnamefont {Kimble}}, \ and\ \bibinfo {author} {\bibfnamefont
			{P.}~\bibnamefont {Zoller}},\ }\bibfield  {title} {\enquote {\bibinfo {title}
			{Cavity opto-mechanics using an optically levitated nanosphere},}\ }\href
	{\doibase 10.1073/pnas.0912969107} {\bibfield  {journal} {\bibinfo  {journal}
			{Proc. Natl. Acad. Sci.}\ }\textbf {\bibinfo {volume} {107}},\ \bibinfo
		{pages} {1005--} (\bibinfo {year} {2010})}\BibitemShut {NoStop}%
	\bibitem [{\citenamefont {Vanner}\ \emph {et~al.}(2015)\citenamefont {Vanner},
		\citenamefont {Pikovski},\ and\ \citenamefont {Kim}}]{Vanner_OM_QST_2015}%
	\BibitemOpen
	\bibfield  {author} {\bibinfo {author} {\bibfnamefont {M.~R.}\ \bibnamefont
			{Vanner}}, \bibinfo {author} {\bibfnamefont {I.}~\bibnamefont {Pikovski}}, \
		and\ \bibinfo {author} {\bibfnamefont {M.~S.}\ \bibnamefont {Kim}},\
	}\bibfield  {title} {\enquote {\bibinfo {title} {Towards optomechanical
				quantum state reconstruction of mechanical motion}},\ }\href {\doibase
		10.1002/andp.201400124} {\bibfield  {journal} {\bibinfo  {journal} {Ann.
				Phys.}\ }\textbf {\bibinfo {volume} {527}},\ \bibinfo {pages} {15} (\bibinfo
		{year} {2015})}\BibitemShut {NoStop}%
	\bibitem [{\citenamefont {Wallentowitz}\ and\ \citenamefont
		{Vogel}(1995)}]{Wallentowitz_QST_ion_1995}%
	\BibitemOpen
	\bibfield  {author} {\bibinfo {author} {\bibfnamefont {S.}~\bibnamefont
			{Wallentowitz}}\ and\ \bibinfo {author} {\bibfnamefont {W.}~\bibnamefont
			{Vogel}},\ }\bibfield  {title} {\enquote {\bibinfo {title} {Reconstruction of
				the quantum mechanical state of a trapped ion}},\ }\href {\doibase
		10.1103/PhysRevLett.75.2932} {\bibfield  {journal} {\bibinfo  {journal}
			{Phys. Rev. Lett.}\ }\textbf {\bibinfo {volume} {75}},\ \bibinfo {pages}
		{2932} (\bibinfo {year} {1995})}\BibitemShut {NoStop}%
	\bibitem [{\citenamefont {Singh}\ and\ \citenamefont
		{Meystre}(2010)}]{Singh_WignerTomo_usingAtom_2010}%
	\BibitemOpen
	\bibfield  {author} {\bibinfo {author} {\bibfnamefont {S.}~\bibnamefont
			{Singh}}\ and\ \bibinfo {author} {\bibfnamefont {P.}~\bibnamefont
			{Meystre}},\ }\bibfield  {title} {\enquote {\bibinfo {title} {Atomic probe
				wigner tomography of a nanomechanical system}},\ }\href {\doibase
		10.1103/PhysRevA.81.041804} {\bibfield  {journal} {\bibinfo  {journal} {Phys.
				Rev. A}\ }\textbf {\bibinfo {volume} {81}},\ \bibinfo {pages} {041804 (R)}
		(\bibinfo {year} {2010})}\BibitemShut {NoStop}%
	\bibitem [{\citenamefont {Parkins}\ and\ \citenamefont
		{Kimble}(1999)}]{Parkins_QST_stateTransfer_1999}%
	\BibitemOpen
	\bibfield  {author} {\bibinfo {author} {\bibfnamefont {A.~S.}\ \bibnamefont
			{Parkins}}\ and\ \bibinfo {author} {\bibfnamefont {H.~J.}\ \bibnamefont
			{Kimble}},\ }\bibfield  {title} {\enquote {\bibinfo {title} {Quantum state
				transfer between motion and light}},\ }\href {\doibase
		10.1088/1464-4266/1/4/323} {\bibfield  {journal} {\bibinfo  {journal} {J.
				Opt.}\ }\textbf {\bibinfo {volume} {1}},\ \bibinfo {pages} {496} (\bibinfo
		{year} {1999})}\BibitemShut {NoStop}%
	\bibitem [{\citenamefont {Wallentowitz}\ and\ \citenamefont
		{Vogel}(1996)}]{Wallentwoitz_QSTunbalancedHomodyne_1996}%
	\BibitemOpen
	\bibfield  {author} {\bibinfo {author} {\bibfnamefont {S.}~\bibnamefont
			{Wallentowitz}}\ and\ \bibinfo {author} {\bibfnamefont {W.}~\bibnamefont
			{Vogel}},\ }\bibfield  {title} {\enquote {\bibinfo {title} {Unbalanced
				homodyning for quantum state measurements}},\ }\href {\doibase
		10.1103/PhysRevA.53.4528} {\bibfield  {journal} {\bibinfo  {journal} {Phys.
				Rev. A}\ }\textbf {\bibinfo {volume} {53}},\ \bibinfo {pages} {4528}
		(\bibinfo {year} {1996})}\BibitemShut {NoStop}%
	\bibitem [{\citenamefont {Banaszek}\ and\ \citenamefont
		{W\'odkiewicz}(1996)}]{Banaszek_QSTphotonCountin_1996}%
	\BibitemOpen
	\bibfield  {author} {\bibinfo {author} {\bibfnamefont {K.}~\bibnamefont
			{Banaszek}}\ and\ \bibinfo {author} {\bibfnamefont {K.}~\bibnamefont
			{W\'odkiewicz}},\ }\bibfield  {title} {\enquote {\bibinfo {title} {Direct
				probing of quantum phase space by photon counting}},\ }\href {\doibase
		10.1103/PhysRevLett.76.4344} {\bibfield  {journal} {\bibinfo  {journal}
			{Phys. Rev. Lett.}\ }\textbf {\bibinfo {volume} {76}},\ \bibinfo {pages}
		{4344} (\bibinfo {year} {1996})}\BibitemShut {NoStop}%
	\bibitem [{\citenamefont {Poyatos}\ \emph {et~al.}(1996)\citenamefont
		{Poyatos}, \citenamefont {Walser}, \citenamefont {Cirac}, \citenamefont
		{Zoller},\ and\ \citenamefont {Blatt}}]{Poyatos_QST_ion_1996}%
	\BibitemOpen
	\bibfield  {author} {\bibinfo {author} {\bibfnamefont {J.~F.}\ \bibnamefont
			{Poyatos}}, \bibinfo {author} {\bibfnamefont {R.}~\bibnamefont {Walser}},
		\bibinfo {author} {\bibfnamefont {J.~I.}\ \bibnamefont {Cirac}}, \bibinfo
		{author} {\bibfnamefont {P.}~\bibnamefont {Zoller}}, \ and\ \bibinfo {author}
		{\bibfnamefont {R.}~\bibnamefont {Blatt}},\ }\bibfield  {title} {\enquote
		{\bibinfo {title} {Motion tomography of a single trapped ion}},\ }\href
	{\doibase 10.1103/PhysRevA.53.R1966} {\bibfield  {journal} {\bibinfo
			{journal} {Phys. Rev. A}\ }\textbf {\bibinfo {volume} {53}},\ \bibinfo
		{pages} {R1966 (R)} (\bibinfo {year} {1996})}\BibitemShut {NoStop}%
	\bibitem [{\citenamefont {Leibfried}\ \emph {et~al.}(1996)\citenamefont
		{Leibfried}, \citenamefont {Meekhof}, \citenamefont {King}, \citenamefont
		{Monroe}, \citenamefont {Itano},\ and\ \citenamefont
		{Wineland}}]{Leibfried_expQSTion_1996}%
	\BibitemOpen
	\bibfield  {author} {\bibinfo {author} {\bibfnamefont {D.}~\bibnamefont
			{Leibfried}}, \bibinfo {author} {\bibfnamefont {D.~M.}\ \bibnamefont
			{Meekhof}}, \bibinfo {author} {\bibfnamefont {B.~E.}\ \bibnamefont {King}},
		\bibinfo {author} {\bibfnamefont {C.}~\bibnamefont {Monroe}}, \bibinfo
		{author} {\bibfnamefont {W.~M.}\ \bibnamefont {Itano}}, \ and\ \bibinfo
		{author} {\bibfnamefont {D.~J.}\ \bibnamefont {Wineland}},\ }\bibfield
	{title} {\enquote {\bibinfo {title} {Experimental determination of the
				motional quantum state of a trapped atom}},\ }\href {\doibase
		10.1103/PhysRevLett.77.4281} {\bibfield  {journal} {\bibinfo  {journal}
			{Phys. Rev. Lett.}\ }\textbf {\bibinfo {volume} {77}},\ \bibinfo {pages}
		{4281} (\bibinfo {year} {1996})}\BibitemShut {NoStop}%
	\bibitem [{\citenamefont {Vanner}\ \emph {et~al.}(2013)\citenamefont {Vanner},
		\citenamefont {Hofer}, \citenamefont {Cole},\ and\ \citenamefont
		{Aspelmeyer}}]{Vanner_CoolingByMsmt_QST_2013}%
	\BibitemOpen
	\bibfield  {author} {\bibinfo {author} {\bibfnamefont {M.~R.}\ \bibnamefont
			{Vanner}}, \bibinfo {author} {\bibfnamefont {J.}~\bibnamefont {Hofer}},
		\bibinfo {author} {\bibfnamefont {G.~D.}\ \bibnamefont {Cole}}, \ and\
		\bibinfo {author} {\bibfnamefont {M.}~\bibnamefont {Aspelmeyer}},\ }\bibfield
	{title} {\enquote {\bibinfo {title} {Cooling-by-measurement and mechanical
				state tomography via pulsed optomechanics}},\ }\href {\doibase
		10.1038/ncomms3295} {\bibfield  {journal} {\bibinfo  {journal} {Nat.
				Commun.}\ }\textbf {\bibinfo {volume} {4}},\ \bibinfo {pages} {2295}
		(\bibinfo {year} {2013})}\BibitemShut {NoStop}%
	\bibitem [{\citenamefont {Romero-Isart}\ \emph
		{et~al.}(2011{\natexlab{a}})\citenamefont {Romero-Isart}, \citenamefont
		{Pflanzer}, \citenamefont {Juan}, \citenamefont {Quidant}, \citenamefont
		{Kiesel}, \citenamefont {Aspelmeyer},\ and\ \citenamefont
		{Cirac}}]{Romero-Isart_LevTheoryProtocols_2011}%
	\BibitemOpen
	\bibfield  {author} {\bibinfo {author} {\bibfnamefont {O.}~\bibnamefont
			{Romero-Isart}}, \bibinfo {author} {\bibfnamefont {A.~C.}\ \bibnamefont
			{Pflanzer}}, \bibinfo {author} {\bibfnamefont {M.~L.}\ \bibnamefont {Juan}},
		\bibinfo {author} {\bibfnamefont {R.}~\bibnamefont {Quidant}}, \bibinfo
		{author} {\bibfnamefont {N.}~\bibnamefont {Kiesel}}, \bibinfo {author}
		{\bibfnamefont {M.}~\bibnamefont {Aspelmeyer}}, \ and\ \bibinfo {author}
		{\bibfnamefont {J.~I.}\ \bibnamefont {Cirac}},\ }\bibfield  {title} {\enquote
		{\bibinfo {title} {Optically levitating dielectrics in the quantum regime:
				Theory and protocols}},\ }\href {\doibase 10.1103/PhysRevA.83.013803}
	{\bibfield  {journal} {\bibinfo  {journal} {Phys. Rev. A}\ }\textbf {\bibinfo
			{volume} {83}},\ \bibinfo {pages} {013803} (\bibinfo {year}
		{2011}{\natexlab{a}})}\BibitemShut {NoStop}%
	\bibitem [{\citenamefont
		{Romero-Isart}(2011)}]{Romero_Isart_QuantumSup_Collapse_2011}%
	\BibitemOpen
	\bibfield  {author} {\bibinfo {author} {\bibfnamefont {O.}~\bibnamefont
			{Romero-Isart}},\ }\bibfield  {title} {\enquote {\bibinfo {title} {Quantum
				superposition of massive objects and collapse models}},\ }\href {\doibase
		10.1103/PhysRevA.84.052121} {\bibfield  {journal} {\bibinfo  {journal} {Phys.
				Rev. A}\ }\textbf {\bibinfo {volume} {84}},\ \bibinfo {pages} {052121}
		(\bibinfo {year} {2011})}\BibitemShut {NoStop}%
	\bibitem [{\citenamefont {Joos}\ \emph {et~al.}(2003)\citenamefont {Joos},
		\citenamefont {Zeh}, \citenamefont {Kiefer}, \citenamefont {Giulini},
		\citenamefont {Kupsch},\ and\ \citenamefont
		{Smatescu}}]{Joos_book_ClassicalWorld_QuTheory_2003}%
	\BibitemOpen
	\bibfield  {author} {\bibinfo {author} {\bibfnamefont {E.}~\bibnamefont
			{Joos}}, \bibinfo {author} {\bibfnamefont {H.~D.}\ \bibnamefont {Zeh}},
		\bibinfo {author} {\bibfnamefont {C.}~\bibnamefont {Kiefer}}, \bibinfo
		{author} {\bibfnamefont {D.}~\bibnamefont {Giulini}}, \bibinfo {author}
		{\bibfnamefont {J.}~\bibnamefont {Kupsch}}, \ and\ \bibinfo {author}
		{\bibfnamefont {I.-O.}\ \bibnamefont {Smatescu}},\ }\href@noop {} {\emph
		{\bibinfo {title} {Decoherence and the Appearance of a Classical World in
				Quantum Theory}}}\ (\bibinfo  {publisher} {New York: Springer},\ \bibinfo
	{year} {2003})\BibitemShut {NoStop}%
	\bibitem [{\citenamefont {Schlosshauer}(2007)}]{Schlosshauer_book_QtoCl_2007}%
	\BibitemOpen
	\bibfield  {author} {\bibinfo {author} {\bibfnamefont {M.~A.}\ \bibnamefont
			{Schlosshauer}},\ }\href@noop {} {\emph {\bibinfo {title} {Decoherence and
				the Quantum-to-Classical Transition}}}\ (\bibinfo  {publisher} {Berlin:
		Springer},\ \bibinfo {year} {2007})\BibitemShut {NoStop}%
	\bibitem [{\citenamefont {Johansson}\ \emph {et~al.}(2012)\citenamefont
		{Johansson}, \citenamefont {Nation},\ and\ \citenamefont
		{Nori}}]{Johansson_qutip_2012}%
	\BibitemOpen
	\bibfield  {author} {\bibinfo {author} {\bibfnamefont {J.~R.}\ \bibnamefont
			{Johansson}}, \bibinfo {author} {\bibfnamefont {P.~D.}\ \bibnamefont
			{Nation}}, \ and\ \bibinfo {author} {\bibfnamefont {F.}~\bibnamefont
			{Nori}},\ }\bibfield  {title} {\enquote {\bibinfo {title} {Qutip: An
				open-source python framework for the dynamics of open quantum systems}},\
	}\href {\doibase 10.1016/j.cpc.2012.02.021} {\bibfield  {journal} {\bibinfo
			{journal} {Comp. Phys. Comm.}\ }\textbf {\bibinfo {volume} {183}},\ \bibinfo
		{pages} {1760} (\bibinfo {year} {2012})}\BibitemShut {NoStop}%
	\bibitem [{\citenamefont {Johansson}\ \emph {et~al.}(2013)\citenamefont
		{Johansson}, \citenamefont {Nation},\ and\ \citenamefont
		{Nori}}]{Johansson_qutip_2013}%
	\BibitemOpen
	\bibfield  {author} {\bibinfo {author} {\bibfnamefont {J.~R.}\ \bibnamefont
			{Johansson}}, \bibinfo {author} {\bibfnamefont {P.~D.}\ \bibnamefont
			{Nation}}, \ and\ \bibinfo {author} {\bibfnamefont {F.}~\bibnamefont
			{Nori}},\ }\bibfield  {title} {\enquote {\bibinfo {title} {Qutip 2: A python
				framework for the dynamics of open quantum systems}},\ }\href {\doibase
		10.1016/j.cpc.2012.11.019} {\bibfield  {journal} {\bibinfo  {journal} {Comp.
				Phys. Comm.}\ }\textbf {\bibinfo {volume} {184}},\ \bibinfo {pages} {1234}
		(\bibinfo {year} {2013})}\BibitemShut {NoStop}%
	\bibitem [{\citenamefont {Ballentine}\ and\ \citenamefont
		{McRae}(1998)}]{Ballentine_MomentEquations_1998}%
	\BibitemOpen
	\bibfield  {author} {\bibinfo {author} {\bibfnamefont {L.~E.}\ \bibnamefont
			{Ballentine}}\ and\ \bibinfo {author} {\bibfnamefont {S.~M.}\ \bibnamefont
			{McRae}},\ }\bibfield  {title} {\enquote {\bibinfo {title} {Moment equations
				for probability distributions in classical and quantum mechanics}},\ }\href
	{\doibase 10.1103/PhysRevA.58.1799} {\bibfield  {journal} {\bibinfo
			{journal} {Phys. Rev. A}\ }\textbf {\bibinfo {volume} {58}},\ \bibinfo
		{pages} {1799} (\bibinfo {year} {1998})}\BibitemShut {NoStop}%
	\bibitem [{\citenamefont {Brizuela}(2014{\natexlab{a}})}]{Brizuela_PRD_2014}%
	\BibitemOpen
	\bibfield  {author} {\bibinfo {author} {\bibfnamefont {D.}~\bibnamefont
			{Brizuela}},\ }\bibfield  {title} {\enquote {\bibinfo {title} {Classical and
				quantum behavior of the harmonic and the quartic oscillators}},\ }\href
	{\doibase 10.1103/PhysRevD.90.125018} {\bibfield  {journal} {\bibinfo
			{journal} {Phys. Rev. D}\ }\textbf {\bibinfo {volume} {90}},\ \bibinfo
		{pages} {125018} (\bibinfo {year} {2014}{\natexlab{a}})}\BibitemShut
	{NoStop}%
	\bibitem [{\citenamefont
		{Brizuela}(2014{\natexlab{b}})}]{Brizuela_PRD_generalizedUncertainty_2014}%
	\BibitemOpen
	\bibfield  {author} {\bibinfo {author} {\bibfnamefont {D.}~\bibnamefont
			{Brizuela}},\ }\bibfield  {title} {\enquote {\bibinfo {title} {Statistical
				moments for classical and quantum dynamics: Formalism and generalized
				uncertainty relations}},\ }\href {\doibase 10.1103/PhysRevD.90.085027}
	{\bibfield  {journal} {\bibinfo  {journal} {Phys. Rev. D}\ }\textbf {\bibinfo
			{volume} {90}},\ \bibinfo {pages} {085027} (\bibinfo {year}
		{2014}{\natexlab{b}})}\BibitemShut {NoStop}%
	\bibitem [{\citenamefont {Torlai}\ \emph {et~al.}(2018)\citenamefont {Torlai},
		\citenamefont {Mazzola}, \citenamefont {Carrasquilla}, \citenamefont
		{Troyer}, \citenamefont {Melko},\ and\ \citenamefont
		{Carleo}}]{Torlai_NN_QST_2018}%
	\BibitemOpen
	\bibfield  {author} {\bibinfo {author} {\bibfnamefont {G.}~\bibnamefont
			{Torlai}}, \bibinfo {author} {\bibfnamefont {G.}~\bibnamefont {Mazzola}},
		\bibinfo {author} {\bibfnamefont {J.}~\bibnamefont {Carrasquilla}}, \bibinfo
		{author} {\bibfnamefont {M.}~\bibnamefont {Troyer}}, \bibinfo {author}
		{\bibfnamefont {R.}~\bibnamefont {Melko}}, \ and\ \bibinfo {author}
		{\bibfnamefont {G.}~\bibnamefont {Carleo}},\ }\bibfield  {title} {\enquote
		{\bibinfo {title} {Neural-network quantum state tomography}},\ }\href
	{\doibase 10.1038/s41567-018-0048-5} {\bibfield  {journal} {\bibinfo
			{journal} {Nat. Phys.}\ }\textbf {\bibinfo {volume} {14}},\ \bibinfo {pages}
		{447} (\bibinfo {year} {2018})}\BibitemShut {NoStop}%
	\bibitem [{\citenamefont {Xu}\ and\ \citenamefont
		{Xu}(2018)}]{Xu_NNforQST_2018}%
	\BibitemOpen
	\bibfield  {author} {\bibinfo {author} {\bibfnamefont {Q.}~\bibnamefont
			{Xu}}\ and\ \bibinfo {author} {\bibfnamefont {S.}~\bibnamefont {Xu}},\
	}\bibfield  {title} {\enquote {\bibinfo {title} {Neural network state
				estimation for full quantum state tomography}},\ }\href
	{https://arxiv.org/abs/1811.06654} {\bibfield  {journal} {\bibinfo  {journal}
			{arXiv:1811.06654}\ } (\bibinfo {year} {2018})}\BibitemShut {NoStop}%
	\bibitem [{\citenamefont {Xin}\ \emph {et~al.}(2018)\citenamefont {Xin},
		\citenamefont {Lu}, \citenamefont {Cao}, \citenamefont {Anikeeva},
		\citenamefont {Lu}, \citenamefont {Li}, \citenamefont {Long},\ and\
		\citenamefont {Zeng}}]{Xin_LocalMsmt_QST_NN_2018}%
	\BibitemOpen
	\bibfield  {author} {\bibinfo {author} {\bibfnamefont {T.}~\bibnamefont
			{Xin}}, \bibinfo {author} {\bibfnamefont {S.}~\bibnamefont {Lu}}, \bibinfo
		{author} {\bibfnamefont {N.}~\bibnamefont {Cao}}, \bibinfo {author}
		{\bibfnamefont {G.}~\bibnamefont {Anikeeva}}, \bibinfo {author}
		{\bibfnamefont {D.}~\bibnamefont {Lu}}, \bibinfo {author} {\bibfnamefont
			{J.}~\bibnamefont {Li}}, \bibinfo {author} {\bibfnamefont {G.}~\bibnamefont
			{Long}}, \ and\ \bibinfo {author} {\bibfnamefont {B.}~\bibnamefont {Zeng}},\
	}\bibfield  {title} {\enquote {\bibinfo {title} {Local-measurement-based
				quantum state tomography via neural networks}},\ }\href
	{https://arxiv.org/abs/1807.07445} {\bibfield  {journal} {\bibinfo  {journal}
			{arXiv:1807.07445}\ } (\bibinfo {year} {2018})}\BibitemShut {NoStop}%
	\bibitem [{\citenamefont {Quek}\ \emph {et~al.}(2018)\citenamefont {Quek},
		\citenamefont {Fort},\ and\ \citenamefont {Ng}}]{Quek_adaptiveQST_NN_2018}%
	\BibitemOpen
	\bibfield  {author} {\bibinfo {author} {\bibfnamefont {Y.}~\bibnamefont
			{Quek}}, \bibinfo {author} {\bibfnamefont {S.}~\bibnamefont {Fort}}, \ and\
		\bibinfo {author} {\bibfnamefont {H.~K.}\ \bibnamefont {Ng}},\ }\bibfield
	{title} {\enquote {\bibinfo {title} {Adaptive quantum state tomography with
				neural networks}},\ }\href {https://arxiv.org/abs/1812.06693} {\bibfield
		{journal} {\bibinfo  {journal} {arXiv:1812.06693}\ } (\bibinfo {year}
		{2018})}\BibitemShut {NoStop}%
	\bibitem [{\citenamefont {Palmieri}\ \emph {et~al.}(2019)\citenamefont
		{Palmieri}, \citenamefont {Kovlakov}, \citenamefont {Bianchi}, \citenamefont
		{Yudin}, \citenamefont {Straupe}, \citenamefont {Biamonte},\ and\
		\citenamefont {Kulik}}]{Palmieri_experimentalQSTenhanced_NN_2019}%
	\BibitemOpen
	\bibfield  {author} {\bibinfo {author} {\bibfnamefont {A.~M.}\ \bibnamefont
			{Palmieri}}, \bibinfo {author} {\bibfnamefont {E.}~\bibnamefont {Kovlakov}},
		\bibinfo {author} {\bibfnamefont {F.}~\bibnamefont {Bianchi}}, \bibinfo
		{author} {\bibfnamefont {D.}~\bibnamefont {Yudin}}, \bibinfo {author}
		{\bibfnamefont {S.}~\bibnamefont {Straupe}}, \bibinfo {author} {\bibfnamefont
			{J.}~\bibnamefont {Biamonte}}, \ and\ \bibinfo {author} {\bibfnamefont
			{S.}~\bibnamefont {Kulik}},\ }\bibfield  {title} {\enquote {\bibinfo {title}
			{Experimental neural network enhanced quantum tomography}},\ }\href
	{https://arxiv.org/abs/1904.05902} {\bibfield  {journal} {\bibinfo  {journal}
			{arXiv:1904.05902}\ } (\bibinfo {year} {2019})}\BibitemShut {NoStop}%
					\bibitem [{\citenamefont {Tebbenjohanns}\ \emph {et~al.}(2019)\citenamefont {Tebbenjohanns},
			\citenamefont {Frimmer},\ and\ \citenamefont
			{Novotny}}]{Tebbenjohanns_optimalDetection_2019}%
		\BibitemOpen
		\bibfield  {author} {\bibinfo {author} {\bibfnamefont {F.}~\bibnamefont
				{Tebbenjohanns}}, \bibinfo {author} {\bibfnamefont {M.}~\bibnamefont
				{Frimmer}},\ and\
			\bibinfo {author} {\bibfnamefont {L.}~\bibnamefont {Novotny}},\ }\bibfield
		{title} {\enquote {\bibinfo {title} {Optimal position detection of a dipolar scatterer in a focused field}},\ }\href {https://journals.aps.org/pra/abstract/10.1103/PhysRevA.100.043821} {\bibfield  {journal} {\bibinfo  {journal}
				{Phys. Rev. A}\ }\textbf {\bibinfo {volume} {100}},\ \bibinfo {pages}
			{043821} (\bibinfo {year} {2019})}\BibitemShut {NoStop}%
		\bibitem [{\citenamefont {Tebbenjohanns}\ \emph {et~al.}(2018)\citenamefont {Tebbenjohanns},
			\citenamefont {Frimmer}, \citenamefont {Jain}, \citenamefont
			{Windey},\ and\ \citenamefont {Novotny}}]{Tebbenjohanns_sidebandAsym_2019}%
		\BibitemOpen
		\bibfield  {author} {\bibinfo {author} {\bibfnamefont {F.}~\bibnamefont
				{Tebbenjohanns}}, \bibinfo {author} {\bibfnamefont {M.}~\bibnamefont {Frimmer}},
			\bibinfo {author} {\bibfnamefont {V.}~\bibnamefont {Jain}}, \bibinfo
			{author} {\bibfnamefont {D.}\ \bibnamefont {Windey}}, \ and\ \bibinfo
			{author} {\bibfnamefont {L.}~\bibnamefont {Novotny}},\ }\bibfield
		{title} {\enquote {\bibinfo {title} {Motional sideband asymmetry of a nanoparticle optically levitated in free space}},\ }\href {https://arxiv.org/abs/1908.05079}
		{\bibfield  {journal} {\bibinfo  {journal} {arXiv:1908.05079}\ } (\bibinfo {year}
			{2019})}\BibitemShut {NoStop}%
	\bibitem [{\citenamefont {Windey}\ \emph {et~al.}(2019)\citenamefont {Windey},
		\citenamefont {Gonzalez-Ballestero}, \citenamefont {Maurer}, \citenamefont
		{Novotny}, \citenamefont {Romero-Isart},\ and\ \citenamefont
		{Reimann}}]{Windey_Cooling_2019}%
		\BibitemOpen
	\bibfield  {author} {\bibinfo {author} {\bibfnamefont {D.}~\bibnamefont
			{Windey}}, \bibinfo {author} {\bibfnamefont {C.}~\bibnamefont
			{Gonzalez-Ballestero}}, \bibinfo {author} {\bibfnamefont {P.}~\bibnamefont
			{Maurer}}, \bibinfo {author} {\bibfnamefont {L.}~\bibnamefont {Novotny}},
		\bibinfo {author} {\bibfnamefont {O.}~\bibnamefont {Romero-Isart}}, \ and\
		\bibinfo {author} {\bibfnamefont {R.}~\bibnamefont {Reimann}},\ }\bibfield
	{title} {\enquote {\bibinfo {title} {Cavity-based 3d cooling of a levitated
				nanoparticle via coherent scattering}},\ }\href {\doibase
		10.1103/PhysRevLett.122.123601} {\bibfield  {journal} {\bibinfo  {journal}
			{Phys. Rev. Lett.}\ }\textbf {\bibinfo {volume} {122}},\ \bibinfo {pages}
		{123601} (\bibinfo {year} {2019})}\BibitemShut {NoStop}%
	\bibitem [{\citenamefont {Gonzalez-Ballestero}\ \emph
		{et~al.}(2019)\citenamefont {Gonzalez-Ballestero}, \citenamefont {Maurer},
		\citenamefont {Windey}, \citenamefont {Novotny}, \citenamefont {Reimann},\
		and\ \citenamefont {Romero-Isart}}]{Gonzalez-Ballestero_CoolingTheoery_2019}%
	\BibitemOpen
	\bibfield  {author} {\bibinfo {author} {\bibfnamefont {C.}~\bibnamefont
			{Gonzalez-Ballestero}}, \bibinfo {author} {\bibfnamefont {P.}~\bibnamefont
			{Maurer}}, \bibinfo {author} {\bibfnamefont {D.}~\bibnamefont {Windey}},
		\bibinfo {author} {\bibfnamefont {L.}~\bibnamefont {Novotny}}, \bibinfo
		{author} {\bibfnamefont {R.}~\bibnamefont {Reimann}}, \ and\ \bibinfo
		{author} {\bibfnamefont {O.}~\bibnamefont {Romero-Isart}},\ }\bibfield
	{title} {\enquote {\bibinfo {title} {Theory for cavity cooling of levitated
				nanoparticles via coherent scattering: Master equation approach}},\ }\href
	{https://journals.aps.org/pra/abstract/10.1103/PhysRevA.100.013805} {\bibfield  {journal} {\bibinfo  {journal}
			{Phys. Rev. A}\ } \textbf {\bibinfo {volume} {100}},\ \bibinfo
		{pages} {013805} (\bibinfo {year} {2019})}\BibitemShut {NoStop}%
	\bibitem [{\citenamefont {Home}\ \emph {et~al.}(2011)\citenamefont
	{Home}, \citenamefont {Hanneke}, \citenamefont {Jost}, \citenamefont {Leibfried},\ and\ \citenamefont
	{Wineland}}]{Home_ionsAnharmonicPot_2011}%
\BibitemOpen
\bibfield  {author} {\bibinfo {author} {\bibfnamefont {J.~P.}~\bibnamefont
		{Home}}, \bibinfo {author} {\bibfnamefont {D.}\ \bibnamefont
		{Hanneke}}, \bibinfo {author} {\bibfnamefont {J.~D.}~\bibnamefont {Jost}}, \bibinfo {author} {\bibfnamefont {D.}~\bibnamefont {Leibfried}} \ and\
	\bibinfo {author} {\bibfnamefont {D.~J.}~\bibnamefont {Wineland}},\ }\bibfield
{title} {\enquote {\bibinfo {title} {Normal modes of trapped ions in the presence of anharmonic trap potentials}},\ }\href {https://doi.org/10.1088/1367-2630/13/7/073026} {\bibfield  {journal} {\bibinfo
		{journal} {New J. Phys.}\ }\textbf {\bibinfo {volume} {13}},\ \bibinfo
	{pages} {073026} (\bibinfo {year} {2011})}\BibitemShut {NoStop}%
	\bibitem [{\citenamefont {Kuhlicke}\ \emph {et~al.}(2014)\citenamefont
		{Kuhlicke}, \citenamefont {Schell}, \citenamefont {Zoll},\ and\ \citenamefont
		{Benson}}]{Kuhlicke_NV_quadrupoleTrap_2014}%
	\BibitemOpen
	\bibfield  {author} {\bibinfo {author} {\bibfnamefont {A.}~\bibnamefont
			{Kuhlicke}}, \bibinfo {author} {\bibfnamefont {A.~W.}\ \bibnamefont
			{Schell}}, \bibinfo {author} {\bibfnamefont {J.}~\bibnamefont {Zoll}}, \ and\
		\bibinfo {author} {\bibfnamefont {O.}~\bibnamefont {Benson}},\ }\bibfield
	{title} {\enquote {\bibinfo {title} {Nitrogen vacancy center fluorescence
				from a submicron diamond cluster levitated in a linear quadrupole ion
				trap}},\ }\href {\doibase 10.1063/1.4893575} {\bibfield  {journal} {\bibinfo
			{journal} {Appl. Phys. Lett.}\ }\textbf {\bibinfo {volume} {105}},\ \bibinfo
		{pages} {073101} (\bibinfo {year} {2014})}\BibitemShut {NoStop}%
	\bibitem [{\citenamefont {Alda}\ \emph {et~al.}(2016)\citenamefont {Alda},
		\citenamefont {Berthelot}, \citenamefont {Rica},\ and\ \citenamefont
		{Quidant}}]{Alda_NanoparticlePaulTrap_2016}%
	\BibitemOpen
	\bibfield  {author} {\bibinfo {author} {\bibfnamefont {I.}~\bibnamefont
			{Alda}}, \bibinfo {author} {\bibfnamefont {J.}~\bibnamefont {Berthelot}},
		\bibinfo {author} {\bibfnamefont {R.~A.}\ \bibnamefont {Rica}}, \ and\
		\bibinfo {author} {\bibfnamefont {R.}~\bibnamefont {Quidant}},\ }\bibfield
	{title} {\enquote {\bibinfo {title} {Trapping and manipulation of individual
				nanoparticles in a planar paul trap}},\ }\href {\doibase 10.1063/1.4965859}
	{\bibfield  {journal} {\bibinfo  {journal} {Appl. Phys. Lett.}\ }\textbf
		{\bibinfo {volume} {109}},\ \bibinfo {pages} {163105} (\bibinfo {year}
		{2016})}\BibitemShut {NoStop}%
	\bibitem [{\citenamefont {Delord}\ \emph {et~al.}(2018)\citenamefont {Delord},
		\citenamefont {Huillery}, \citenamefont {Schwab}, \citenamefont {Nicolas},
		\citenamefont {Lecordier},\ and\ \citenamefont
		{H\'etet}}]{Delord_SpinEchosLevi_2018}%
	\BibitemOpen
	\bibfield  {author} {\bibinfo {author} {\bibfnamefont {T.}~\bibnamefont
			{Delord}}, \bibinfo {author} {\bibfnamefont {P.}~\bibnamefont {Huillery}},
		\bibinfo {author} {\bibfnamefont {L.}~\bibnamefont {Schwab}}, \bibinfo
		{author} {\bibfnamefont {L.}~\bibnamefont {Nicolas}}, \bibinfo {author}
		{\bibfnamefont {L.}~\bibnamefont {Lecordier}}, \ and\ \bibinfo {author}
		{\bibfnamefont {G.}~\bibnamefont {H\'etet}},\ }\bibfield  {title} {\enquote
		{\bibinfo {title} {Ramsey interferences and spin echoes from electron spins
				inside a levitating macroscopic particle}},\ }\href {\doibase
		10.1103/PhysRevLett.121.053602} {\bibfield  {journal} {\bibinfo  {journal}
			{Phys. Rev. Lett.}\ }\textbf {\bibinfo {volume} {121}},\ \bibinfo {pages}
		{053602} (\bibinfo {year} {2018})}\BibitemShut {NoStop}%
	\bibitem [{\citenamefont {Bykov}\ \emph {et~al.}(2019)\citenamefont {Bykov},
		\citenamefont {Mestres}, \citenamefont {Dania}, \citenamefont {Schm\"oger},\
		and\ \citenamefont {Northup}}]{Bykov_NanoparticlesPaulTrap_2019}%
	\BibitemOpen
	\bibfield  {author} {\bibinfo {author} {\bibfnamefont {D.~S.}\ \bibnamefont
			{Bykov}}, \bibinfo {author} {\bibfnamefont {P.}~\bibnamefont {Mestres}},
		\bibinfo {author} {\bibfnamefont {L.}~\bibnamefont {Dania}}, \bibinfo
		{author} {\bibfnamefont {L.}~\bibnamefont {Schm\"oger}}, \ and\ \bibinfo
		{author} {\bibfnamefont {T.~E.}\ \bibnamefont {Northup}},\ }\bibfield
	{title} {\enquote {\bibinfo {title} {Direct loading of nanoparticles under
				high vacuum into a Paul trap for levitodynamical experiments}},\ }\href
	{https://aip.scitation.org/doi/10.1063/1.5109645} {\bibfield  {journal} {\bibinfo  {journal}
			{Appl. Phys. Lett}\ }\textbf {\bibinfo {volume} {115}},\ \bibinfo {pages}
		{034101} (\bibinfo {year} {2019})}\BibitemShut {NoStop}%
		\bibitem [{\citenamefont {Romero-Isart}\ \emph {et~al.}(2012)\citenamefont
		{Romero-Isart}, \citenamefont {Clemente}, \citenamefont {Navau},
		\citenamefont {Sanchez},\ and\ \citenamefont
		{Cirac}}]{Romero-Isart_MagneticLevi_2012}%
	\BibitemOpen
	\bibfield  {author} {\bibinfo {author} {\bibfnamefont {O.}~\bibnamefont
			{Romero-Isart}}, \bibinfo {author} {\bibfnamefont {L.}~\bibnamefont
			{Clemente}}, \bibinfo {author} {\bibfnamefont {C.}~\bibnamefont {Navau}},
		\bibinfo {author} {\bibfnamefont {A.}~\bibnamefont {Sanchez}}, \ and\
		\bibinfo {author} {\bibfnamefont {J.~I.}\ \bibnamefont {Cirac}},\ }\bibfield
	{title} {\enquote {\bibinfo {title} {Quantum magnetomechanics with levitating
				superconducting microspheres}},\ }\href {\doibase
		10.1103/PhysRevLett.109.147205} {\bibfield  {journal} {\bibinfo  {journal}
			{Phys. Rev. Lett.}\ }\textbf {\bibinfo {volume} {109}},\ \bibinfo {pages}
		{147205} (\bibinfo {year} {2012})}\BibitemShut {NoStop}%
	\bibitem [{\citenamefont {Slezak}\ \emph {et~al.}(2018)\citenamefont {Slezak},
		\citenamefont {Lewandowski}, \citenamefont {Hsu},\ and\ \citenamefont
		{D'Urso}}]{Slezak_MagneticLevi_2018}%
	\BibitemOpen
	\bibfield  {author} {\bibinfo {author} {\bibfnamefont {B.~R.}\ \bibnamefont
			{Slezak}}, \bibinfo {author} {\bibfnamefont {C.~W.}\ \bibnamefont
			{Lewandowski}}, \bibinfo {author} {\bibfnamefont {J.-F.}\ \bibnamefont
			{Hsu}}, \ and\ \bibinfo {author} {\bibfnamefont {B.}~\bibnamefont {D'Urso}},\
	}\bibfield  {title} {\enquote {\bibinfo {title} {Cooling the motion of a
				silica microsphere in a magneto-gravitational trap in ultra-high vacuum}},\
	}\href {\doibase 10.1088/1367-2630/aacac1} {\bibfield  {journal} {\bibinfo
			{journal} {New J. Phys.}\ }\textbf {\bibinfo {volume} {20}},\ \bibinfo
		{pages} {063028} (\bibinfo {year} {2018})}\BibitemShut {NoStop}%
	\bibitem [{\citenamefont {Prat-Camps}\ \emph {et~al.}(2017)\citenamefont
		{Prat-Camps}, \citenamefont {Teo}, \citenamefont {Rusconi}, \citenamefont
		{Wieczorek},\ and\ \citenamefont
		{Romero-Isart}}]{Prat-Camps_InertialForceSensor_2017}%
	\BibitemOpen
	\bibfield  {author} {\bibinfo {author} {\bibfnamefont {J.}~\bibnamefont
			{Prat-Camps}}, \bibinfo {author} {\bibfnamefont {C.}~\bibnamefont {Teo}},
		\bibinfo {author} {\bibfnamefont {C.~C.}\ \bibnamefont {Rusconi}}, \bibinfo
		{author} {\bibfnamefont {W.}~\bibnamefont {Wieczorek}}, \ and\ \bibinfo
		{author} {\bibfnamefont {O.}~\bibnamefont {Romero-Isart}},\ }\bibfield
	{title} {\enquote {\bibinfo {title} {Ultrasensitive inertial and force
				sensors with diamagnetically levitated magnets}},\ }\href {\doibase
		10.1103/PhysRevApplied.8.034002} {\bibfield  {journal} {\bibinfo  {journal}
			{Phys. Rev. Appl.}\ }\textbf {\bibinfo {volume} {8}},\ \bibinfo {pages}
		{034002} (\bibinfo {year} {2017})}\BibitemShut {NoStop}%
	\bibitem [{\citenamefont {Romero-Isart}\ \emph
		{et~al.}(2011{\natexlab{b}})\citenamefont {Romero-Isart}, \citenamefont
		{Pflanzer}, \citenamefont {Blaser}, \citenamefont {Kaltenbaek}, \citenamefont
		{Kiesel}, \citenamefont {Aspelmeyer},\ and\ \citenamefont
		{Cirac}}]{Romero-Isart_LargeQuSup_2011}%
	\BibitemOpen
	\bibfield  {author} {\bibinfo {author} {\bibfnamefont {O.}~\bibnamefont
			{Romero-Isart}}, \bibinfo {author} {\bibfnamefont {A.~C.}\ \bibnamefont
			{Pflanzer}}, \bibinfo {author} {\bibfnamefont {F.}~\bibnamefont {Blaser}},
		\bibinfo {author} {\bibfnamefont {R.}~\bibnamefont {Kaltenbaek}}, \bibinfo
		{author} {\bibfnamefont {N.}~\bibnamefont {Kiesel}}, \bibinfo {author}
		{\bibfnamefont {M.}~\bibnamefont {Aspelmeyer}}, \ and\ \bibinfo {author}
		{\bibfnamefont {J.~I.}\ \bibnamefont {Cirac}},\ }\bibfield  {title} {\enquote
		{\bibinfo {title} {Large quantum superpositions and interference of massive
				nanometer-sized objects}},\ }\href {\doibase 10.1103/PhysRevLett.107.020405}
	{\bibfield  {journal} {\bibinfo  {journal} {Phys. Rev. Lett.}\ }\textbf
		{\bibinfo {volume} {107}},\ \bibinfo {pages} {020405} (\bibinfo {year}
		{2011}{\natexlab{b}})}\BibitemShut {NoStop}%
	\bibitem [{\citenamefont {Hebestreit}\ \emph {et~al.}(2018)\citenamefont
		{Hebestreit}, \citenamefont {Frimmer}, \citenamefont {Reimann},\ and\
		\citenamefont {Novotny}}]{Hebestreit_FreeFallingNanoparticles_2018}%
	\BibitemOpen
	\bibfield  {author} {\bibinfo {author} {\bibfnamefont {E.}~\bibnamefont
			{Hebestreit}}, \bibinfo {author} {\bibfnamefont {M.}~\bibnamefont {Frimmer}},
		\bibinfo {author} {\bibfnamefont {R.}~\bibnamefont {Reimann}}, \ and\
		\bibinfo {author} {\bibfnamefont {L.}~\bibnamefont {Novotny}},\ }\bibfield
	{title} {\enquote {\bibinfo {title} {Sensing static forces with free-falling
				nanoparticles}},\ }\href {\doibase 10.1103/PhysRevLett.121.063602} {\bibfield
		{journal} {\bibinfo  {journal} {Phys. Rev. Lett.}\ }\textbf {\bibinfo
			{volume} {121}},\ \bibinfo {pages} {063602} (\bibinfo {year}
		{2018})}\BibitemShut {NoStop}%
	\bibitem [{\citenamefont {Diehl}\ \emph {et~al.}(2018)\citenamefont {Diehl},
		\citenamefont {Hebestreit}, \citenamefont {Reimann}, \citenamefont
		{Tebbenjohanns}, \citenamefont {Frimmer},\ and\ \citenamefont
		{Novotny}}]{Diehl_NanoparticleCloseToMembrane_2018}%
	\BibitemOpen
	\bibfield  {author} {\bibinfo {author} {\bibfnamefont {R.}~\bibnamefont
			{Diehl}}, \bibinfo {author} {\bibfnamefont {E.}~\bibnamefont {Hebestreit}},
		\bibinfo {author} {\bibfnamefont {R.}~\bibnamefont {Reimann}}, \bibinfo
		{author} {\bibfnamefont {F.}~\bibnamefont {Tebbenjohanns}}, \bibinfo {author}
		{\bibfnamefont {M.}~\bibnamefont {Frimmer}}, \ and\ \bibinfo {author}
		{\bibfnamefont {L.}~\bibnamefont {Novotny}},\ }\bibfield  {title} {\enquote
		{\bibinfo {title} {Optical levitation and feedback cooling of a nanoparticle
				at subwavelength distances from a membrane}},\ }\href {\doibase
		10.1103/PhysRevA.98.013851} {\bibfield  {journal} {\bibinfo  {journal} {Phys.
				Rev. A}\ }\textbf {\bibinfo {volume} {98}},\ \bibinfo {pages} {013851}
		(\bibinfo {year} {2018})}\BibitemShut {NoStop}%
	\bibitem [{\citenamefont {Magrini}\ \emph {et~al.}(2018)\citenamefont
		{Magrini}, \citenamefont {Norte}, \citenamefont {Riedinger}, \citenamefont
		{Marinkovi\'{c}}, \citenamefont {Grass}, \citenamefont {Deli\'{c}},
		\citenamefont {Gr\"{o}blacher}, \citenamefont {Hong},\ and\ \citenamefont
		{Aspelmeyer}}]{Magrini_NearFieldNanoparticle_2018}%
	\BibitemOpen
	\bibfield  {author} {\bibinfo {author} {\bibfnamefont {L.}~\bibnamefont
			{Magrini}}, \bibinfo {author} {\bibfnamefont {R.~A.}\ \bibnamefont {Norte}},
		\bibinfo {author} {\bibfnamefont {R.}~\bibnamefont {Riedinger}}, \bibinfo
		{author} {\bibfnamefont {I.}~\bibnamefont {Marinkovi\'{c}}}, \bibinfo
		{author} {\bibfnamefont {D.}~\bibnamefont {Grass}}, \bibinfo {author}
		{\bibfnamefont {U.}~\bibnamefont {Deli\'{c}}}, \bibinfo {author}
		{\bibfnamefont {S.}~\bibnamefont {Gr\"{o}blacher}}, \bibinfo {author}
		{\bibfnamefont {S.}~\bibnamefont {Hong}}, \ and\ \bibinfo {author}
		{\bibfnamefont {M}~\bibnamefont {Aspelmeyer}},\ }\bibfield  {title} {\enquote
		{\bibinfo {title} {Near-field coupling of a levitated nanoparticle to a
				photonic crystal cavity}},\ }\href {\doibase 10.1364/OPTICA.5.001597}
	{\bibfield  {journal} {\bibinfo  {journal} {Optica}\ }\textbf {\bibinfo
			{volume} {5}},\ \bibinfo {pages} {1597} (\bibinfo {year} {2018})}\BibitemShut
	{NoStop}%
	\bibitem [{\citenamefont {Pino}\ \emph {et~al.}(2018)\citenamefont {Pino},
		\citenamefont {Prat-Camps}, \citenamefont {Sinha}, \citenamefont
		{Venkatesh},\ and\ \citenamefont {Romero-Isart}}]{Pino_Skatepark_2018}%
	\BibitemOpen
	\bibfield  {author} {\bibinfo {author} {\bibfnamefont {H.}~\bibnamefont
			{Pino}}, \bibinfo {author} {\bibfnamefont {J.}~\bibnamefont {Prat-Camps}},
		\bibinfo {author} {\bibfnamefont {K.}~\bibnamefont {Sinha}}, \bibinfo
		{author} {\bibfnamefont {B.~P.}\ \bibnamefont {Venkatesh}}, \ and\ \bibinfo
		{author} {\bibfnamefont {O.}~\bibnamefont {Romero-Isart}},\ }\bibfield
	{title} {\enquote {\bibinfo {title} {On-chip quantum interference of a
				superconducting microsphere}},\ }\href {\doibase 10.1088/2058-9565/aa9d15}
	{\bibfield  {journal} {\bibinfo  {journal} {Quantum. Sci. Technol.}\ }\textbf
		{\bibinfo {volume} {3}},\ \bibinfo {pages} {025001} (\bibinfo {year}
		{2018})}\BibitemShut {NoStop}%
		\bibitem [{\citenamefont {Romero-Isart} (2017)\citenamefont {Romero-Isart}}]{ORI_CoherentInflation_2017}%
	\BibitemOpen
	\bibfield  {author} {\bibinfo {author} {\bibfnamefont {O.}~\bibnamefont
			{Romero-Isart}},\ }\bibfield
	{title} {\enquote {\bibinfo {title} {Coherent inflation for large quantum superpositions of levitated microspheres}},\ }\href {https://doi.org/10.1088/1367-2630/aa99bf}
	{\bibfield  {journal} {\bibinfo  {journal} {New J. Phys.}\ }\textbf
		{\bibinfo {volume} {19}},\ \bibinfo {pages} {123029} (\bibinfo {year}
		{2017})}\BibitemShut {NoStop}%
	\bibitem [{\citenamefont {Deli\ifmmode~\acute{c}\else \'{c}\fi{}}\ \emph
		{et~al.}(2019)\citenamefont {Deli\ifmmode~\acute{c}\else \'{c}\fi{}},
		\citenamefont {Reisenbauer}, \citenamefont {Grass}, \citenamefont {Kiesel},
		\citenamefont {Vuleti\ifmmode~\acute{c}\else \'{c}\fi{}},\ and\ \citenamefont
		{Aspelmeyer}}]{Delic_Cooling_2019}%
	\BibitemOpen
	\bibfield  {author} {\bibinfo {author} {\bibfnamefont {U.}~\bibnamefont
			{Deli\ifmmode~\acute{c}\else \'{c}\fi{}}}, \bibinfo {author} {\bibfnamefont
			{M.}~\bibnamefont {Reisenbauer}}, \bibinfo {author} {\bibfnamefont
			{D.}~\bibnamefont {Grass}}, \bibinfo {author} {\bibfnamefont
			{N.}~\bibnamefont {Kiesel}}, \bibinfo {author} {\bibfnamefont
			{V.}~\bibnamefont {Vuleti\ifmmode~\acute{c}\else \'{c}\fi{}}}, \ and\
		\bibinfo {author} {\bibfnamefont {M.}~\bibnamefont {Aspelmeyer}},\ }\bibfield
	{title} {\enquote {\bibinfo {title} {Cavity cooling of a levitated
				nanosphere by coherent scattering}},\ }\href {\doibase
		10.1103/PhysRevLett.122.123602} {\bibfield  {journal} {\bibinfo  {journal}
			{Phys. Rev. Lett.}\ }\textbf {\bibinfo {volume} {122}},\ \bibinfo {pages}
		{123602} (\bibinfo {year} {2019})}\BibitemShut {NoStop}%
	\bibitem [{\citenamefont {Jain}\ \emph {et~al.}(2016)\citenamefont {Jain},
		\citenamefont {Gieseler}, \citenamefont {Moritz}, \citenamefont {Dellago},
		\citenamefont {Quidant},\ and\ \citenamefont
		{Novotny}}]{Jain_LeviRecoilHeating_2016}%
	\BibitemOpen
	\bibfield  {author} {\bibinfo {author} {\bibfnamefont {V.}~\bibnamefont
			{Jain}}, \bibinfo {author} {\bibfnamefont {J.}~\bibnamefont {Gieseler}},
		\bibinfo {author} {\bibfnamefont {C.}~\bibnamefont {Moritz}}, \bibinfo
		{author} {\bibfnamefont {C.}~\bibnamefont {Dellago}}, \bibinfo {author}
		{\bibfnamefont {R.}~\bibnamefont {Quidant}}, \ and\ \bibinfo {author}
		{\bibfnamefont {L.}~\bibnamefont {Novotny}},\ }\bibfield  {title} {\enquote
		{\bibinfo {title} {Direct measurement of photon recoil from a levitated
				nanoparticle}},\ }\href {\doibase 10.1103/PhysRevLett.116.243601} {\bibfield
		{journal} {\bibinfo  {journal} {Phys. Rev. Lett.}\ }\textbf {\bibinfo
			{volume} {116}},\ \bibinfo {pages} {243601} (\bibinfo {year}
		{2016})}\BibitemShut {NoStop}%
	\bibitem [{\citenamefont {Fonseca}\ \emph {et~al.}(2016)\citenamefont
		{Fonseca}, \citenamefont {Aranas}, \citenamefont {Millen}, \citenamefont
		{Monteiro},\ and\ \citenamefont
		{Barker}}]{Fonesca_NonlinearDynNanopart_2016}%
	\BibitemOpen
	\bibfield  {author} {\bibinfo {author} {\bibfnamefont {P.~Z.~G.}\
			\bibnamefont {Fonseca}}, \bibinfo {author} {\bibfnamefont {E.~B.}\
			\bibnamefont {Aranas}}, \bibinfo {author} {\bibfnamefont {J.}~\bibnamefont
			{Millen}}, \bibinfo {author} {\bibfnamefont {T.~S.}\ \bibnamefont
			{Monteiro}}, \ and\ \bibinfo {author} {\bibfnamefont {P.~F.}\ \bibnamefont
			{Barker}},\ }\bibfield  {title} {\enquote {\bibinfo {title} {Nonlinear
				dynamics and strong cavity cooling of levitated nanoparticles}},\ }\href
	{\doibase 10.1103/PhysRevLett.117.173602} {\bibfield  {journal} {\bibinfo
			{journal} {Phys. Rev. Lett.}\ }\textbf {\bibinfo {volume} {117}},\ \bibinfo
		{pages} {173602} (\bibinfo {year} {2016})}\BibitemShut {NoStop}%
	\bibitem [{\citenamefont {Millen}\ \emph {et~al.}(2015)\citenamefont {Millen},
		\citenamefont {Fonseca}, \citenamefont {Mavrogordatos}, \citenamefont
		{Monteiro},\ and\ \citenamefont
		{Barker}}]{Millen_CoolingChargedNanosphere_2015}%
	\BibitemOpen
	\bibfield  {author} {\bibinfo {author} {\bibfnamefont {J.}~\bibnamefont
			{Millen}}, \bibinfo {author} {\bibfnamefont {P.~Z.~G.}\ \bibnamefont
			{Fonseca}}, \bibinfo {author} {\bibfnamefont {T.}~\bibnamefont
			{Mavrogordatos}}, \bibinfo {author} {\bibfnamefont {T.~S.}\ \bibnamefont
			{Monteiro}}, \ and\ \bibinfo {author} {\bibfnamefont {P.~F.}\ \bibnamefont
			{Barker}},\ }\bibfield  {title} {\enquote {\bibinfo {title} {Cavity cooling a
				single charged levitated nanosphere}},\ }\href {\doibase
		10.1103/PhysRevLett.114.123602} {\bibfield  {journal} {\bibinfo  {journal}
			{Phys. Rev. Lett.}\ }\textbf {\bibinfo {volume} {114}},\ \bibinfo {pages}
		{123602} (\bibinfo {year} {2015})}\BibitemShut {NoStop}%
	\bibitem [{\citenamefont {Asenbaum}\ \emph {et~al.}(2013)\citenamefont
		{Asenbaum}, \citenamefont {Kuhn}, \citenamefont {Nimmrichter}, \citenamefont
		{Sezer},\ and\ \citenamefont {Arndt}}]{Asenbaum_CoolingSiliconNanopart_2013}%
	\BibitemOpen
	\bibfield  {author} {\bibinfo {author} {\bibfnamefont {P.}~\bibnamefont
			{Asenbaum}}, \bibinfo {author} {\bibfnamefont {S.}~\bibnamefont {Kuhn}},
		\bibinfo {author} {\bibfnamefont {S.}~\bibnamefont {Nimmrichter}}, \bibinfo
		{author} {\bibfnamefont {U.}~\bibnamefont {Sezer}}, \ and\ \bibinfo {author}
		{\bibfnamefont {M.}~\bibnamefont {Arndt}},\ }\bibfield  {title} {\enquote
		{\bibinfo {title} {Cavity cooling of free silicon nanoparticles in high
				vacuum}},\ }\href {\doibase 10.1038/ncomms3743} {\bibfield  {journal}
		{\bibinfo  {journal} {Nat. Commun.}\ }\textbf {\bibinfo {volume} {4}},\
		\bibinfo {pages} {2743} (\bibinfo {year} {2013})}\BibitemShut {NoStop}%
	\bibitem [{\citenamefont {Kiesel}\ \emph {et~al.}(2013)\citenamefont {Kiesel},
		\citenamefont {Blaser}, \citenamefont {Deli{\'c}}, \citenamefont {Grass},
		\citenamefont {Kaltenbaek},\ and\ \citenamefont
		{Aspelmeyer}}]{Kiesel_CoolingSubmicronPart_2013}%
	\BibitemOpen
	\bibfield  {author} {\bibinfo {author} {\bibfnamefont {N.}~\bibnamefont
			{Kiesel}}, \bibinfo {author} {\bibfnamefont {F.}~\bibnamefont {Blaser}},
		\bibinfo {author} {\bibfnamefont {U.}~\bibnamefont {Deli{\'c}}}, \bibinfo
		{author} {\bibfnamefont {D.}~\bibnamefont {Grass}}, \bibinfo {author}
		{\bibfnamefont {R.}~\bibnamefont {Kaltenbaek}}, \ and\ \bibinfo {author}
		{\bibfnamefont {M.}~\bibnamefont {Aspelmeyer}},\ }\bibfield  {title}
	{\enquote {\bibinfo {title} {Cavity cooling of an optically levitated
				submicron particle}},\ }\href {\doibase 10.1073/pnas.1309167110} {\bibfield
		{journal} {\bibinfo  {journal} {Proc. Natl. Acad. Sci.}\ }\textbf {\bibinfo
			{volume} {110}},\ \bibinfo {pages} {14180--14185} (\bibinfo {year}
		{2013})}\BibitemShut {NoStop}%
	\bibitem [{\citenamefont {Gieseler}\ \emph {et~al.}(2012)\citenamefont
		{Gieseler}, \citenamefont {Deutsch}, \citenamefont {Quidant},\ and\
		\citenamefont {Novotny}}]{Gieseler_FeedbackCoolingNanopart_2012}%
	\BibitemOpen
	\bibfield  {author} {\bibinfo {author} {\bibfnamefont {J.}~\bibnamefont
			{Gieseler}}, \bibinfo {author} {\bibfnamefont {B.}~\bibnamefont {Deutsch}},
		\bibinfo {author} {\bibfnamefont {R.}~\bibnamefont {Quidant}}, \ and\
		\bibinfo {author} {\bibfnamefont {L.}~\bibnamefont {Novotny}},\ }\bibfield
	{title} {\enquote {\bibinfo {title} {Subkelvin parametric feedback cooling of
				a laser-trapped nanoparticle}},\ }\href {\doibase
		10.1103/PhysRevLett.109.103603} {\bibfield  {journal} {\bibinfo  {journal}
			{Phys. Rev. Lett.}\ }\textbf {\bibinfo {volume} {109}},\ \bibinfo {pages}
		{103603} (\bibinfo {year} {2012})}\BibitemShut {NoStop}%
	\bibitem [{\citenamefont {Li}\ \emph {et~al.}(2013)\citenamefont {Li},
		\citenamefont {Kheifets},\ and\ \citenamefont
		{Raizen}}]{Li_CoolingMicrosphere_2011}%
	\BibitemOpen
	\bibfield  {author} {\bibinfo {author} {\bibfnamefont {T.}~\bibnamefont
			{Li}}, \bibinfo {author} {\bibfnamefont {S.}~\bibnamefont {Kheifets}}, \ and\
		\bibinfo {author} {\bibfnamefont {M.~G.}\ \bibnamefont {Raizen}},\ }\bibfield
	{title} {\enquote {\bibinfo {title} {Millikelvin cooling of an optically
				trapped microsphere in vacuum}},\ }\href {\doibase 10.1038/nphys1952}
	{\bibfield  {journal} {\bibinfo  {journal} {Nat. Phys.}\ }\textbf {\bibinfo
			{volume} {7}},\ \bibinfo {pages} {527} (\bibinfo {year} {2013})}\BibitemShut
	{NoStop}%
\end{thebibliography}
\end{document}